\documentclass[acmlarge]{acmart}
\usepackage{natbib}  % or the citation package you prefer
\usepackage{graphicx} % Required for inserting images
\usepackage{hyperref}
\usepackage{soul}
\usepackage{amsmath}
\usepackage{gensymb}

\hypersetup{
    colorlinks=true,
    linkcolor=blue,
    urlcolor=blue,
    citecolor=blue,
}
\usepackage{booktabs}
\usepackage{multirow}
\usepackage{subcaption} 
\usepackage{caption}
\usepackage[export]{adjustbox}
\usepackage{subcaption}

\title{Beyond Questionnaires: Video Analysis for Social Anxiety Detection}

\author{Nilesh Kumar Sahu}
\email{nilesh21@iiserb.ac.in}
\orcid{0000-0003-1675-7270}
\author{Nandigramam Sai Harshit}
\email{nandigramam19@iiserb.ac.in}
\author{Rishabh Uikey}
\email{rishabh18@iiserb.ac.in}
\author{Haroon R. Lone}
\email{haroon@iiserb.ac.in}
\affiliation{%
  \institution{Indian Institute of Science Education and Research, Bhopal}
  \streetaddress{Bhauri}
  \city{Bhopal}
  \state{Madhya Pradesh}
  \country{India}
  \postcode{462066}
}

\begin{document}

\begin{abstract}
Social Anxiety Disorder (SAD) significantly impacts individuals' daily lives and relationships. The conventional methods for SAD detection involve physical consultations and self-reported questionnaires, but they have limitations such as time consumption and bias. 
This paper introduces video analysis as a promising method for early SAD detection. Specifically, we present a new approach for detecting SAD in individuals from various bodily features extracted from the video data. We conducted a study to collect video data of 92 participants performing impromptu speech in a controlled environment. Using the video data, we studied the behavioral change in participants' head, body, eye gaze, and action units. By applying a range of machine learning and deep learning algorithms, we achieved an accuracy rate of up to 74\% in classifying participants as SAD or non-SAD.
Video-based SAD detection offers a non-intrusive and scalable approach that can be deployed in real-time, potentially enhancing early detection and intervention capabilities. 

\end{abstract}



\maketitle

\section{Introduction}
Anxiety disorders encompass excessive fear, worry, and disruptive behaviors, often involving symptoms like nervousness, dread, hyperventilation, trembling, and more. These symptoms can cause significant distress and impair daily functioning. In 2019, an estimated 301 million individuals, including 58 million children and adolescents, were affected by anxiety disorders~\cite{mental_disorders_who}. Various types of anxiety disorders exist, such as generalized anxiety disorder, panic disorder, social anxiety disorder, and separation anxiety disorder. Social anxiety disorder (SAD), also known as social phobia, is a prevalent anxiety disorder impacting individuals regularly. Those with SAD experience intense self-consciousness and embarrassment, driven by the fear of negative judgment or scrutiny by others. These elevated anxiety levels often disrupt relationships, daily routines, and professional activities.

The conventional approach for detecting Social Anxiety Disorder (SAD) involves physical consultations with psychiatrists or psychologists and the use of self-reported questionnaire-style assessments like the Depression Anxiety Stress Scale (DASS)\cite{Dass-cite}, Social Interaction Anxiety Scale (SIAS)\cite{article}, and Social Phobia Inventory (SPIN)~\cite{connor_davidson}. However, these methods have limitations in accurately identifying anxiety in individuals. Physical consultations tend to be time-consuming and are susceptible to human biases, leading to varying diagnoses among different practitioners. Additionally, questionnaire-based assessments are prone to issues such as poor recall and bias, potentially impacting the accuracy of anxiety detection.

Acknowledging the limitations of current diagnostic tools, researchers have begun exploring digital and mobile approaches for screening SAD. These methods encompass diverse measures, including physiological indicators like heart rate, behavioral markers such as ``time spent by SAD individuals in specific locations,'' and the analysis of acoustic features extracted from audio recordings. While these approaches offer non-invasive options, their outcomes are still evolving, prompting ongoing research endeavors.

In addition to these methods, video analysis of participants emerges as a promising avenue for detecting SAD. Video-based SAD detection presents a non-invasive and scalable solution that can be implemented in real-time, offering a versatile approach to identifying SAD across various settings. This innovative approach adds to the array of SAD diagnostic tools, potentially strengthening early detection and intervention capabilities.

In this study, we have used video data of participants to explore the behavioral cues to detect SAD. Our study encompasses many bodily indicators, including body pose, head pose, action units (facial features), and the eye gaze of an individual, in order to classify SAD and non-SAD participants. We used a low-cost (approx \$81) phone camera to capture participant's video during a speech activity. Speech is a well-known stress-inducing activity. It follows the Trier Social Stress Test (TSST), a protocol used to induce moderate psychological stress in a laboratory setting, and evaluated its effects on physiological responses \cite{kirschbaum1993trier}. From the captured videos, we extracted the head pose landmarks, body pose landmarks and their positions in space, facial features (action units), and the eye gaze coordinates in the 3D space of the participants. Following the feature extraction, we used the correlation analysis to identify and drop the redundant features. Further, several machine learning and deep learning classification techniques were used for classifying SAD and non-SAD participants. We achieved an impressive classification accuracy of 74\% deep learning methods. Following are the contributions of our work.

\begin{enumerate}

\item Our research presents a new approach for detecting SAD from various bodily features using machine learning and deep learning methods. Our work is a first-of-its-kind to detect SAD using bodily features collected with a low-end smartphone during an anxiety-provoking speech activity. 

\item We release our processed dataset publicly containing bodily features extracted from the video-recorded data of the participants~\cite{our-video-dataset}. The dataset includes features like eye gaze, head pose, body pose, and action units (facial features).
\end{enumerate}

\section{Related work}

Social dominance and submission are the characteristics of SAD individuals and are conveyed through an array of nonverbal cues, including facial expressions, eye contact, and body language~\cite{galili2013acoustic}. Studies have explored visual features like head motion, facial muscle movement, etc., to screen mental disorders in individuals. However, to the best of our knowledge, no studies or analysis exists where visual features were explored for understanding the behavior of SAD individuals. In this section, we will discuss the existing approaches to understand the behavioral cues of specific mental disorders. 
%\hl{There are psychology paper for understanding the behavior of SAD individuals, however there are papers where the researchers have used ML or DL on it, so we will need to include a literature review on psychology paper, at least a paragraph}

Giannakakis et al.~\cite{giannakakis2017stress} focused on detecting and analyzing stress/anxiety emotional states using facial cues. The participants in the study were exposed to neutral, relaxed, and stressed/anxious states induced by external stressors. At the same time, the participant's facial expressions were captured on video. The study developed a framework to utilize various facial cues, including eye-related features such as gaze and direction and head motion parameters. The study's findings suggest that specific facial cues can be used to distinguish between different stress and anxiety states. Similarly, Sun et al. \cite{sun2022estimating} investigated remote assessment methods for detecting stress. They employed a variety of features, including remote photoplethysmography, action units, eye gaze, and head pose. These features were extracted from facial videos recorded during online video meetings to estimate stress levels in participants. The study demonstrated that combining behavioral and remote signal features yielded higher accuracy in estimating stress levels than using individual features alone.

In a similar kind of study, Horigome et al.~\cite{horigome2020evaluating} videotaped participants' upper bodies using Red-Green-Blue-Depth (RGB-D) sensors during the conversation with the psychologist. They compared the head motion of patients with major depressive disorders (MDD) and bipolar disorders (BD) to the head motion of healthy control (HC) participants. They found that patients with depression moved significantly slower than HC in 5th and 50th percentile motion speed. Further, using the upper body motion data as an input measure and self-reported depression as a label, the authors were able to discriminate between normal and mild, normal and moderate, and moderate and severe with moderate agreement depression severity. 

{\bf Deep learning approaches:} The use of deep learning methods to detect mental disorders has significantly increased. In one of the works by Zhang et al. \cite{zhang2020video}, a two-level stress detection framework called TSDNet was developed using facial expressions and action motions in the video. TSDNet was modeled to learn face-level and action-level representations separately. The results were then fused through a stream-weighted integrator with local and global attention for stress identification. TSDNet outperformed feature engineering approaches with a detection accuracy of 85.42\% and an F1-score of 85.28\%. They also noted that considering both facial and action motions could improve accuracy and F1-Score by more than 7\% compared to the face level-only or action level-only methods. Similarly, Yang et al.~\cite{8465953} utilized linguistic features besides audio and video features to build depression detection systems. The framework consists of three parts: (i) A Deep Convolutional Network and Deep Neural Network, (ii) A paragraph vector and Support Vector Machine model, and (iii) A Random Forest model. The framework improved the accuracies of depression estimation and depression classification, with an F1-measure of up to 0.746 on the depression sub-challenge of AVEC2016.

Zhou et al.~\cite{8344107} developed a deep regression network named `DepressNet'  to learn a depression representation with a visual explanation. The network trains multiple models for various regions of the face, whose results are later fused to improve the overall recognition performance. 
Also, Liu et al.~\cite{liu2022measuring} developed a multi-modal deep convolutional neural network to evaluate the severity of depression symptoms in real time, which was based on facial expressions and body movements of participants captured through ordinary cameras. They found that information from different modes, when appropriately integrated, significantly improved the model's accuracy, resulting in 74\%  similarity in the results and self-rating anxiety scale (SAS) and Hamilton Depression scale (HAMD). 

Gavrilescu et al.~\cite{gavrilescu2019predicting} built a non-intrusive architecture of three layers to analyze facial expressions using the Facial Action Coding System (FACS) to determine the Depression Anxiety Stress Scale (DASS) levels. The first layer used Active Appearance Models (AAM) and a set of multiclass Support Vector Machines (SVM) for Action Unit (AU) classification. The second layer built a matrix containing the AUs' intensity levels. The third layer used an optimal feedforward neural network (FFNN) to analyze the matrix from the second layer in a pattern recognition task, predicting the DASS levels. The architecture was able to discriminate between healthy subjects and those affected by Major Depressive Disorder (MDD) by 93\%, and 85\% in the case of Generalized Anxiety Disorder (GAD).

Zhu et al.~\cite{7812588} developed a new approach to predict the Beck Depression Inventory-II (BDI-II) values from video data based on deep networks. The framework is designed in a two-stream manner, which captures both facial appearance and dynamics. Joint tuning layers integrate this information to improve the results of the analysis. Also, Melo et al.~\cite{8756568} developed a new method to use 3D convolutional networks to capture spatio-temporal information. A 3D convolutional model is used over the entire face of the participant, termed global, and specific regions of the face (e.g., eyes) are termed local. The features extracted from global and local regions are merged, and a global average pooling method is proposed to summarize the spatio-temporal features of the final layer.

{\it Research Gap:} Most previous work in this domain focuses primarily on depression detection through video features. In this paper, we have utilized video data of participants to explore behavioral cues to detect social anxiety disorder. Along with head pose, action units (facial features), and eye gaze, we have also used body pose (shoulders, elbows, wrists) and their summarized features to detect anxiety. When utilizing these features, we can gain a deeper insight into the best features to detect social anxiety disorder.
\begin{comment}
\hl{Haroon: Find relevant papers}
\begin{enumerate}
    \item Giannakakis et al.~\cite{giannakakis2017stress}
    \item Sun et al. \cite{sun2022estimating}
    \item Horigome et al.~\cite{horigome2020evaluating}
    \item Gavrilescu et al.~\cite{gavrilescu2019predicting}
\end{enumerate}
Deep learning ones
\begin{enumerate}
    \item Zhang et al. \cite{zhang2020video}
    \item Yang et al. \cite{8465953}
    \item Zhou et al.~\cite{8344107}
    \item Liu et al.~\cite{liu2022measuring} 
    \item Zhu et al.~\cite{7812588}
    \item Melo et al.~\cite{8756568} 
\end{enumerate}

\textcolor{red}{How is SAD prediction different than depression prediction?} \\
\end{comment}

\section{Data Collection}
%In this section, we will discuss about the various phases of how the study was conducted.
\subsection{Participants recruitment}
After the approval from the Institute’s Review Board, we recruited student participants from our Institute for the study.
 A Google Form was forwarded to all the students via email. The form had two sections: (i) Basic study information, including the study's aim, expectations, and timeline. (ii) A section for collecting participants' demographics and availability from the provided time slots.
Initially, 250 students filled out the Google form and confirmed their participation in the study. However, more than 50\% of these participants later withdrew for  personal reasons, such as ``I am not on campus today'' or ``I am no longer interested.''

For each study session, we emailed the group participants (P1, P2, and P3) again and confirmed the date, time, and venue of the study. We grouped the interested participants into groups of three, ensuring the grouped participants did not know each other prior to the study. 

\begin{figure}
    \centering
    \includegraphics[scale=0.21]{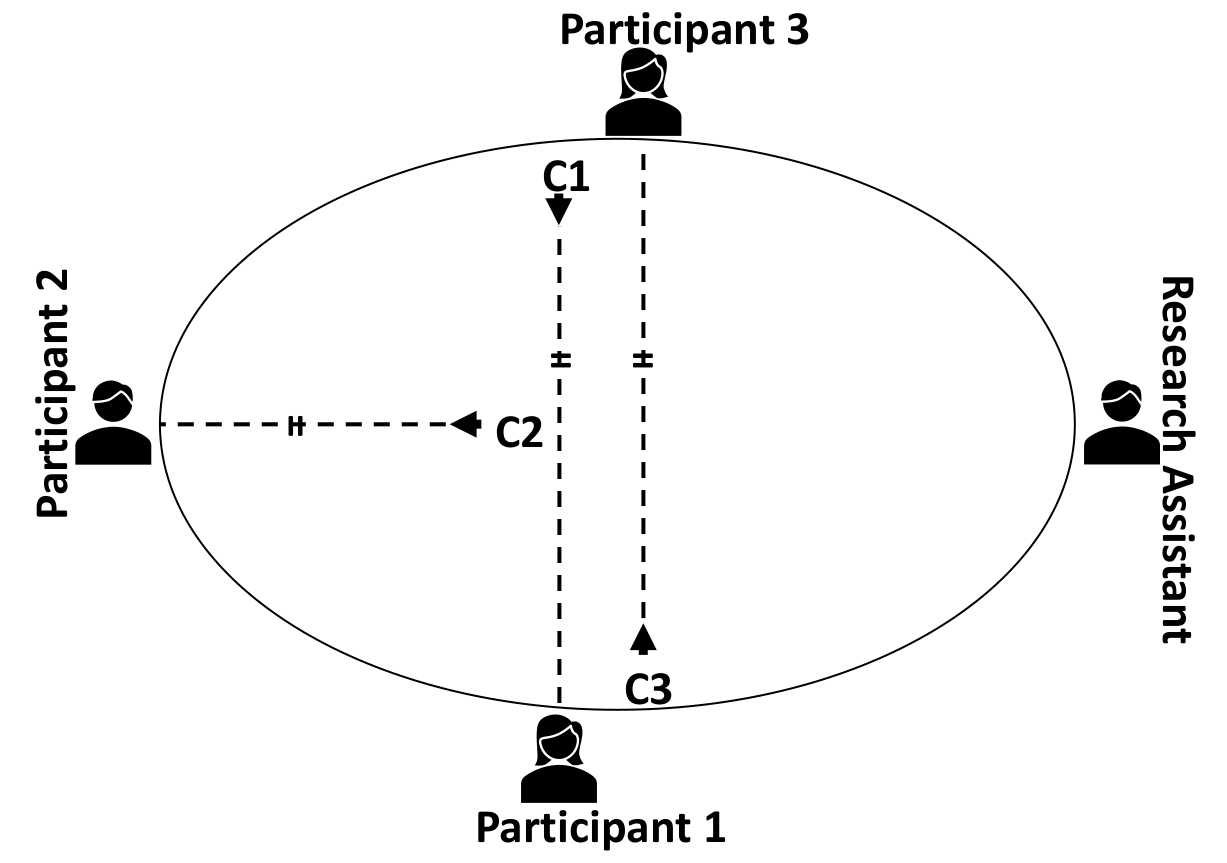}
    \caption{Study setup. C1, C2, and C3 represent cameras 1, 2, and 3, respectively}
    \label{fig:video_setup}
\end{figure}

% \begin{figure}
%   \begin{minipage}[b]{0.47\textwidth}
%     \centering
%     \includegraphics[scale=0.21]{Figures/video_setup_v1.png}
%     \caption{Study setup. C1, C2, and C3 represent camera 1, 2, and 3, respectively}
%     \label{fig:video_setup}
%   \end{minipage}%
%   \begin{minipage}[b]{0.4\textwidth}
%     \centering
%     \includegraphics[scale=0.39]{Figures/plots/distribution.pdf}
%     \caption{Participants' distribution according to self-reported score post activity \hl{bring this picture in sr}}
%     \label{fig: distribution}
%   \end{minipage}
% \end{figure}

\begin{figure}
    \centering
    \includegraphics[scale=0.5]{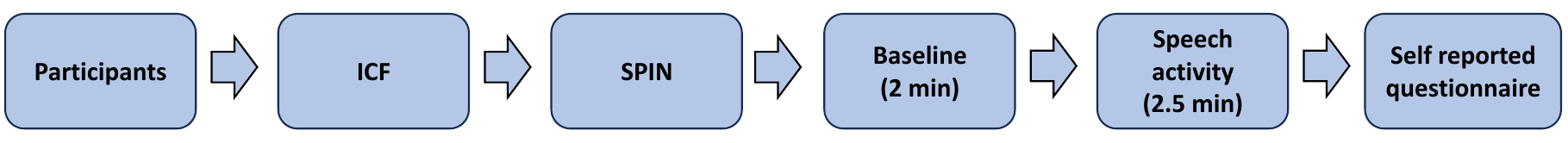}
    \caption{Procedure of the Study}
    \label{fig:StudyFlow}
\end{figure}

On the specified date and time, these participants gathered in one of the Institute's rooms for the study. We had two dummy participants ready if a study participant did not show up. Upon arrival, a research assistant (RA) welcomed them and arranged them around a round table, as shown in Figure~\ref{fig:video_setup}. The RA then explained the study to the participants. After obtaining their signed consent, the participants were asked to fill out the SPIN \cite{connor2000psychometric} form. Figure~\ref{fig:StudyFlow} shows the step-by-step procedure of the study.

% the distribution of participants is shown in Figure~\ref{fig: distribution}\hl{update this later}, and a self-reported social anxiety assessment questionnaire

Next, video recording via smartphones mounted on the table was initiated, as shown in Figure \ref{fig:video_setup}. We used an Android smartphone (Redmi 9A) with a 13-megapixel camera and a video recording application called ``Background Video Recorder''. Three smartphone cameras, C1, C2, and C3 captured participants P1, P2, and P3, respectively, throughout the study. Table \ref{tab: specification} shows the specification of video camera.

First, the participants went through the baseline period, during which they were instructed to sit quietly in their positions for 2 minutes. Next, participants reported their anxiety on a scale from 1 to 5, with 1 indicating minimum anxiety and 5 signifying maximum anxiety. After that, the research assistant (RA) explained the Trier Social Stress Test (TSST) that the participants were required to perform. We selected Speech as the TSST, as it is widely recognized for its ability to induce anxiety and has been utilized in various studies \cite{garcia2017autonomic,harrewijn2018heart}. The speech topics include ``Artificial Intelligence and the problem of unemployment'', ``Agriculture vs manufacturing industry in our country'', etc.

Following the explanation, the RA assigned a speech topic to Participant 1 (P1) and instructed him/her to speak for at least 2.5 minutes, with a 30-second preparation time. After the preparation period, P1 delivered the speech on the assigned topic in front of the RA and two other participants. Post-speech activity, the participant reported the level of anxiety he/she experienced during the speech
in another form (discussed in section \ref{Self-reported Questionnaire}). This process was then repeated for Participant 2 (P2) and Participant 3 (P3), each of whom also had a 30-second preparation period before delivering their speeches. 

After the study session, the captured video data is then transferred to the Lab's server. It took us four months and 40 study sessions to collect data from 111 participants. After removing the missing and noisy data, we were left with 92 participants' data. Table \ref{tab: participant_demo} displays the participants' demographic information of the 92 participants.

\begin{table}[h]
\captionsetup{justification=centering} % Center-align the caption
\captionof{table}{Video camera specifications} % Use \captionof to place caption above the table
\centering % Center-align the table
\begin{tabular}{@{}cc@{}}
\toprule
\textbf{Specification}      & \textbf{Setting}          \\ \midrule
Video Resolution            & 1920x1080 (Full HD 1080p) \\
Recording Bit-Rate          & 6 Mb/s (Full HD 1080p)    \\
Aspect Ratio                & 16:9                      \\
Max frames-per-second (FPS) & 5 to 30                   \\
Zoom                        & 1.0x Digital Zoom         \\ \bottomrule
\end{tabular}
\label{tab: specification}
\end{table}

\begin{table}[h]
\centering
\caption{Participants demographic information. SR - Self-reported State anxiety}
\begin{tabular}{@{}ccccc@{}}
\toprule
 & \textbf{count} & \textbf{SR Score} & \textbf{Age}  & \textbf{Gender}\\
  & \textbf{\#} & \textbf{($\mu$, $\sigma$)} & {($\mu$, $\sigma$)} & {(Male, Female)} \\ \midrule
\textbf{All} & 92 & (14.5, 4.6)& (21, 2.7)& (61, 31)\\
\textbf{Only SAD} & 45& (18.4, 2.7)& (20.8, 2.6)& (29, 16)\\
\textbf{Only Non-SAD} & 47& (10.7, 2.5)& (21.2, 2.8)& (32, 15)\\ \bottomrule
\end{tabular}
\vspace{0.1cm}
\label{tab: participant_demo}
\end{table}

\subsection{Self-reported Questionnaire} \label{Self-reported Questionnaire}
The self-reported questionnaire administered immediately after the speech activity was meticulously designed to assess participants' state anxiety during the speech shown in Table \ref{tab:survey_questions}. It was crafted based on an extensive literature review and consultations with a clinical psychiatrist. To ensure its validity, the questionnaire underwent rigorous validation processes, including face, content, and construct validity assessments. This involved feedback and input from students, counselors, and psychiatrists to refine and validate the questionnaire's effectiveness and accuracy in capturing participants' anxiety levels during the speech activity.

The questionnaire, presented in Table \ref{tab:survey_questions}, covers three distinct aspects of social anxiety: negative evaluation (Q5), scrutiny (Q4), and avoidance tendencies (Q3). Additionally, it includes two additional queries aimed at gauging anxiety levels (Q1) and the participants' perceptions of their performance (Q2). These questions collectively evaluate participants' anxiety levels during the activity. Participants provided responses to each question on a scale ranging from 1 to 5. This scale facilitated the assessment of varying degrees of anxiety  related to their experience during the activity.

\begin{table}[]
\small
\caption{Survey questionnaire collected at the end of the speech activity.}
\label{tab:survey_questions}
\begin{tabular}{lc}
\toprule
\textbf{Question }                                                                                         & \textbf{Response (Scale of 1 to 5)}   \\ \midrule
Q1. I felt extreme nervousness in the last activity.& Strongly Disagree (1) ... Strongly Agree (5) \\
Q2. During the last activity, how did you feel about yourself?& Unhappy/Negative (1) ... Happy/Positive (5)  \\
Q3. I wish I could have avoided the last activity as it produced      anxiety in me.& Strongly Disagree (1) ... Strongly Agree (5) \\
Q4. During the last activity, I was worried about being scrutinized by others.& Strongly Disagree (1) ... Strongly Agree (5) \\
Q5. During the last activity, I thought my behavior was negatively evaluated.& Strongly Disagree (1) ... Strongly Agree (5) \\ \bottomrule
\end{tabular}
\end{table}
 
%\hl{let's understand the distribution of responses post the anxious activity}

\subsection{Data pre-processing}

This section discusses various pre-processing steps taken to prepare the initial dataset. The initial dataset corresponds to the anxiety score and the video data collected during the speech activity.

% This section discusses various pre-processing steps taken to prepare the initial dataset. The initial dataset corresponds to the speech activity and self-reported questionnaire shown in Figure~\ref{fig:StudyFlow}.

\subsubsection{\textbf{Total social anxiety score:}} \label{section_socialanxiety_score_processing}
\begin{figure}
    \centering
    \includegraphics[scale=0.39]{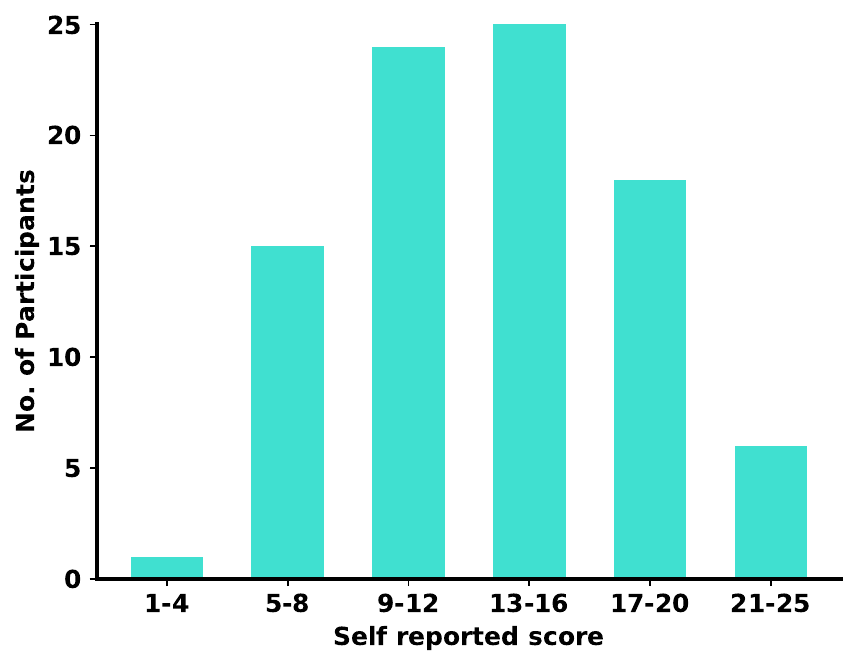}
    \caption{Participants' distribution according to total self-reported social anxiety score.}
    \label{fig: distribution}
\end{figure}

The total social anxiety score for each participant was computed by adding the responses for each question provided in the self-reported questionnaire collected at the end of the speech activity. The total score for a participant ranges from a minimum of 5 to a maximum of 25. Figure~\ref{fig: distribution} shows the distribution of participants based on total self-report anxiety score. Notably, the response for Q2, `During the last activity, how did you feel about yourself?', was reversed, as negative responses were on the left side of the Likert scale (i.e., 1 and 2). Moreover, a higher total score indicates that participants experienced higher anxiety, and a lower total score indicates that participants experienced lower anxiety. Subsequently, the total anxiety scores were utilized to categorize participants into SAD and non-SAD groups. 

We selected a threshold of 15, where participants with a total self-reported response greater than and equal to 15 were categorized as SAD-experiencing participants. At the same time, those with scores below 15 were considered non-SAD-experiencing participants. The rationale for choosing a threshold of 15 stems from its position as the median value between the minimum (5) and maximum (25) scores. Additionally, if a participant provides a response of 3 to all questions, the final anxiety score will be 15, which implies that the participant has not chosen any responses that indicate the non-existence of anxiety. Moreover, setting the threshold at 15 in our dataset results in a nearly balanced representation, with 45 individuals as having SAD and 47 individuals as non-SAD.

\subsubsection{\textbf{Video data:}} 
We clipped and extracted the speech activity from the collected video data for each participant. Although the designated speech activity duration was 2.5 minutes, some participants spoke for more than this duration, while others spoke for less than 2 minutes. Moreover, a few participants either spoke minimally or refrained from performing the speech activity, and the data of such participants were omitted. Furthermore, we used a threshold of 1.5 minutes during video clipping, ensuring homogeneity in the dataset. This pre-processing step led to a reduction in the total number of participant data from 100 to 92. Further, various features were computed using the pre-processed data, as discussed in the below section.

% We clipped and extracted the speech portions from the collected video data for each participant. While reviewing the clipped video data, we observed that some participants had spoken for less than 1 minute. Therefore, to maintain homogeneity in the dataset and to standardize the consensus in the data, we clipped all the participant data to a threshold of 1 minute and 30 seconds. This pre-processing step resulted in a reduction in the total number of participant data to 92. Using This pre-processed data, various features have been computed, which are discussed further.

\section{Feature Extraction}
%\hl{fix my place at the end}
\begin{figure}
    \centering
    \includegraphics[scale=0.5]{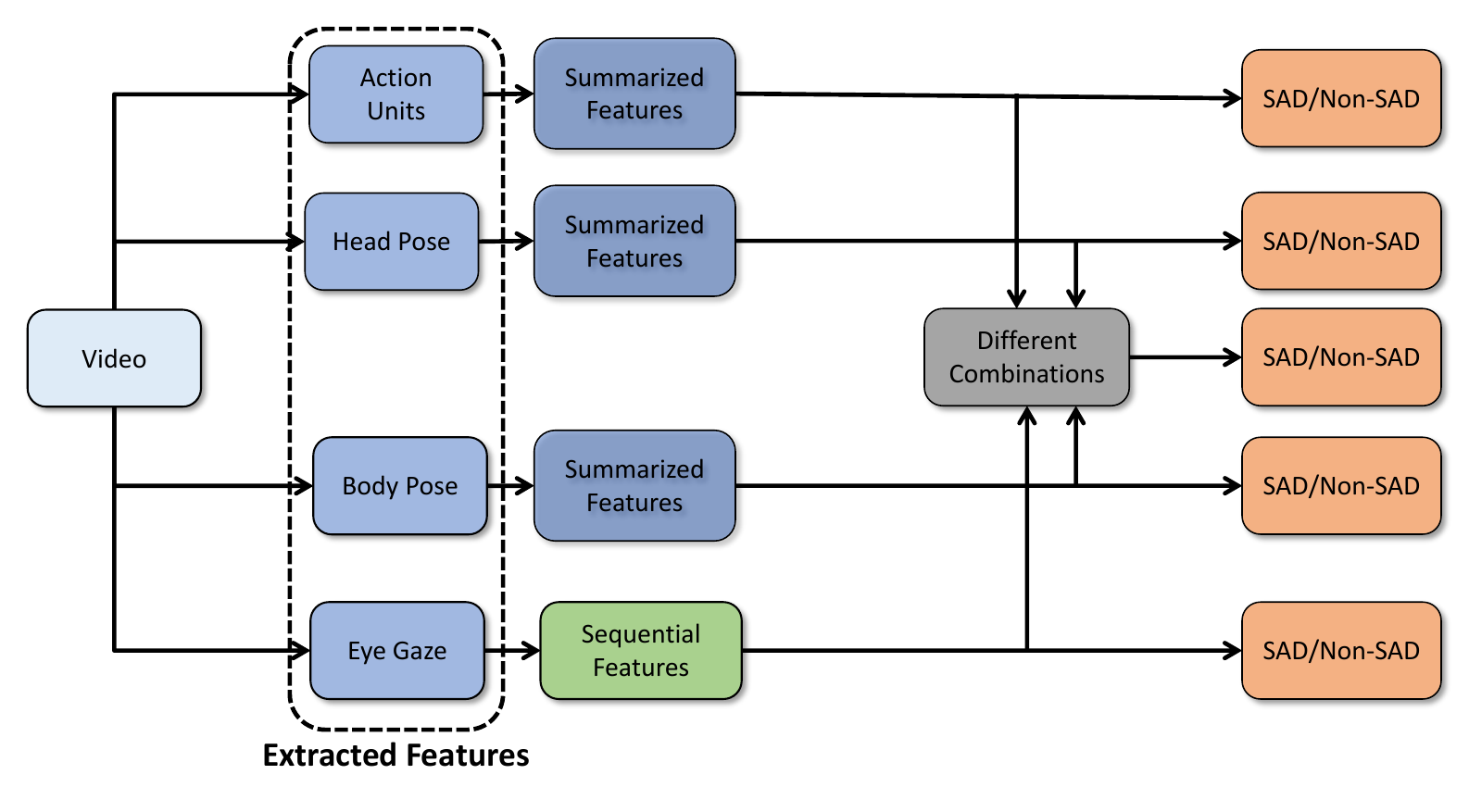}
    \caption{Methodology flowchart}
    \label{fig:methodology-flowchart}
\end{figure}
 
Feature extraction refers to the process of extracting meaningful features from the collected raw data. 
From the processed video data, we extracted features corresponding to (i) head posture, (ii) body posture, (iii) facial features (action units), and (iv) eye gaze. Literature suggests that when people are stressed or anxious, their head and body posture can shift, their facial expressions may display increased tension or exhibit specific action units associated with distress, and their eye gaze may become more erratic or focused on particular cues \cite{galili2013acoustic}. 
These features provide valuable insights into human behavior, especially in situations where individuals experience stress or anxiety~\cite{horigome2020evaluating}.
Moreover, the significance of these features is underscored by works like \cite{sun2022estimating}, where authors also used such features to see the behavioral patterns of individuals under stress. 
We categorize the extracted features as Summarized and Sequential features, as discussed below. Figure~\ref{fig:methodology-flowchart} shows a high-level summary of our methodology. 
%We further divide the extracted features into two categories:- participant-specific features and temporal features.

%\subsubsection{\textbf{Participant-specific features}}
\subsection{ Summarized features}
Summarized features are statistical attributes that are computed without considering the sequential order, as the order does not impact the feature's value. For instance, the mean of a numerical sequence (2, 3, 9) remains the same even when the sequence is different (9, 2, 3), indicating that sequential order is irrelevant when calculating the ``mean'' of numbers. In our analysis, we computed summarized features from  Head Pose, Body Pose, and Action Units as the sequential order of Head, Body, and Face movements during the speech activity does not influence the computation of such features. We extracted these features from every video frame as shown in Figure~\ref{fig:summarized_features}. This section elaborates the methods employed to extract summarized features.

\begin{itemize}

\item \textbf{\textit{Head posture:}}
Head posture refers to the position of the head in a given 3-dimensional space. To calculate the head postures of participants, we utilized the Py-Feat library \cite{cheong2023pyfeat}. The Py-Feat library is a widely recognized package that allows the extraction of head pose metrics for every frame in a given video input. Using this library, we calculated the 3-dimensional spatial coordinates of the head's position, along with the head's pitch (looking up and down), yaw (looking left and right), and roll (tilting head left and right), determined by facial landmarks such as the eyes, nose, and mouth. From these measurements, we converted the values from degrees to Gaussian distances. We computed 72 statistical features representing the head pose for each metric, namely mean, median, mode, minimum, maximum, range, standard deviation, skewness, and kurtosis for each of the 92 participants, resulting in a feature vector of size 92 x 72.

\begin{figure}[!t]
    \centering
        \begin{subfigure}{\textwidth}
        \includegraphics [width=\textwidth]{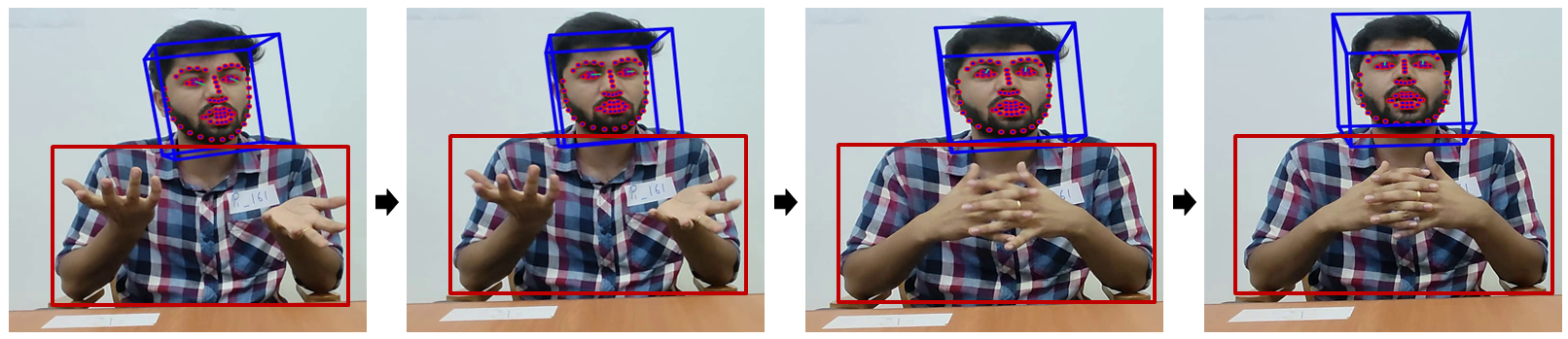}
        \caption{Head pose (Blue colored bounding boxes), Body pose (Red colored bounding boxes), and Action units (Pink colored dots around the face)}
        \label{fig:summarized_features}
    \end{subfigure} 
    \begin{subfigure}{\textwidth}
        \includegraphics [width=\textwidth]{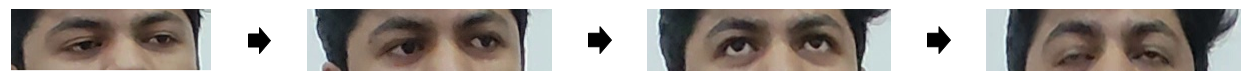}
        \caption{Eye gaze}
        \label{fig:sequential}
    \end{subfigure}

    \caption{Examples showing the extracted features from continuous video frames [{\bf Best viewed in color}]}
\end{figure}

\item \textbf{\textit{Body posture:}}
Body posture refers to the body's orientation and the position of various body parts (e.g., shoulders, arms, hands) relative to each other in space. We only used the body parts above the waist level of the participant since the participants were seated during the study. We utilized the MediaPipe library~\cite{lugaresi2019mediapipe} to derive the 3-dimensional co-ordinates for body posture landmarks (eyes, nose, mouth, shoulders, wrists, elbows, fingers) in each frame of a video, where the video has a frame rate of 27 FPS (Frames Per Second). Using these metrics, we computed velocity, acceleration, and motion coordinates for the key body landmarks for every frame. Subsequently, we calculated statistical features such as mean, median, mode, minimum, maximum, range, standard deviation,  skewness, and kurtosis for each landmark metric for each participant, creating a feature vector consisting of size 92 x 161 for a total of 92 participants.

\item \textbf{\textit{Action Units (Facial Features):}} Action units (AUs) represent fundamental actions of individual muscles or groups of muscles \cite{908962}. To extract action units from participant videos, we employed the OpenFace 2.0 library \cite{8373812}. We chose OpenFace 2.0 due to its widespread use as a facial behavior analysis toolkit, providing comprehensive facial information. This toolkit adheres to the Facial Action Coding System (FACS), assigning a unique action unit to each facial region. OpenFace 2.0 outputs 20 action units, scored on a scale of 0 to 5 for every frame in the input video, where 0 indicates the action unit was not detected, and 5 signifies successful detection.
We computed statistical features for these action units, including mean, median, mode, range, minimum, maximum, standard deviation, skewness, and kurtosis. These statistics were used to generate a feature vector consisting of 119 features for each participant representing the action units, resulting in a feature vector of size 92 x 119.

\end{itemize}

\begin{table}
    \centering
    \tiny
    \caption{List of non-redundant features identified with correlation analysis. These features were used in the data analysis. Acronym, PTP in Action units row refers to peak to peak.}
    \begin{tabular}{lll}
    \toprule
          &Feature & Statistical Measure  \\ \midrule
          && \\
         &Left Eye (X-Direction)  & Change in X-coordinates, rolling mean, rolling standard deviation, change in gaze $(\theta\degree)$ \\
         &Left Eye (Y-Direction) & Change in Y-coordinates, rolling mean, rolling standard deviation, change in gaze $(\theta\degree)$ \\
         &Left Eye (Z-Direction)  & Change in Z-coordinates, rolling mean, rolling standard deviation, change in gaze $(\theta\degree)$ \\
         \textbf{Eye gaze}&Right Eye (X-Direction)  & Change in X-coordinates, rolling mean, rolling standard deviation, change in gaze $(\theta\degree)$   \\
         &Right Eye (Y-Direction) & Change in Y-coordinates, rolling mean, rolling standard deviation, change in gaze $(\theta\degree)$\\
         &Right Eye (Z-Direction) & Change in Z-coordinates, rolling mean, rolling standard deviation, change in gaze $(\theta\degree)$ \\
         &Pitch & Angle $(\theta\degree)$, Change in pitch\\
         &Roll & Angle $(\theta\degree)$, Change in roll\\
         &Yaw & Angle $(\theta\degree)$, Change in angle \\
         &Participant ID & Unique identifier assigned to each participant\\ \midrule
         &AU01 (Inner Brow Raiser)& Median, PTP, Skewness\\
         &AU02 (Outer Brow Raiser)& PTP, Skewness\\
         &AU04 (Brow Lowerer) & Mean, Mode, PTP, Min, Skewness\\
         &AU05 (Upper Lid Raiser) & Mean, PTP, Skewness\\
         &AU06 (Cheek Raiser) & Mean, PTP, Skewness\\
         &AU07 (Lid Tightener) & Mean, Mode, PTP, Skewness\\
         &AU09 (Nose Wrinkler) & Mean, PTP, Skewness\\
         \textbf{Action units}&AU10 (Upper Lip Raiser) & Mean, PTP, Skewness\\
         &AU12 (Lip Corner Puller) &Mean, PTP, Skewness\\
         &AU14 (Dimpler)&Mean, PTP, Skewness\\
         &AU15 (Lip Corner Depressor) &Mean, PTP, Skewness\\
         &AU17 (Chin Raiser) &Mean, PTP, Skewness\\
         &AU20 (Lip Stretcher) &Mean, Skewness\\
         &AU23 (Lip Tightener) &Mean, PTP, Skewness\\
         &AU25 (Lips Part) &Mean, PTP, Skewness\\
         &AU26 (Jaw Drop) &Mean, PTP, Skewness\\
         &AU45 (Blink) &Mean, Median, PTP, Skewness\\ \midrule
         &Left Ear & Velocity, Acceleration, MoveRangeX, MoveRangeY, MoveRangeZ, NetMoveRange \\
         \textbf{Body pose}&Left Elbow & Acceleration, MovetRangeX \\
         &Left Index & Acceleration, MoveRangeX, MoveRangeY\\
         &Right Index & Acceleration\\ \midrule
         &Pitch Velocity     & Mean, Median, Mode, Skewness, Kurtosis\\
         &Roll Velocity      & Mean, Median, Mode, Skewness, Kurtosis\\
         &Yaw Velocity       & Mean, Median, Mode, Skewness\\
         \textbf{Head pose}&Pitch Acceleration & Mean, Median, Skewness\\
         &Roll Acceleration  & Mean, Median, Skewness\\
         &Yaw Acceleration   & Mean, Median, Skewness\\
         &Final Velocity     & Mode\\
         &Final Acceleration & Mode\\ \midrule
    \end{tabular}
    \label{tab:non-redundant-features}
\end{table}

\subsection{Sequential Features} \label{Sequential Features}
Sequential features are computed while accounting for the temporal order of the input data. 
Considering eye gaze data as sequential is crucial as it provides a deeper understanding of participants' behavior varying over time.
 
Eye gaze refers to the orientation (Figure~\ref{fig:sequential}) of the eyes within the face, which can provide information about the direction where an individual is looking in the space. We employed the OpenFace 2.0 library~\cite{8373812} to determine the direction of participants' eye gazes. Later, we derived various features, including pitch, yaw, roll, and the rate of change of gaze direction.
    We further computed statistical measures from the extracted features, such as rolling mean, median, standard deviation, and min-max values, while preserving temporal order. To achieve this, we resampled our participant data and processed the information sequentially to prevent data mixing between participants. Due to resampling, there were instances of missing values (i.e.,{\tt NaN}) in the dataset. To address this issue, we applied different interpolation techniques like zero fill and backfill to fill in all the missing {\tt NaN} values. Additionally, we divided our resampled data into sequences where each sequence is of the same length and treated as a single entity when fed to the classifier. This approach allowed classifiers to capture and interpret changes in temporal behavior effectively.

\subsection{Dataset} \label{section:dataset}

The final dataset, formed by combining head pose, body pose, action units, and eye gaze features, resulted in 383 features for all 92 participants. Individually, the head pose, body pose, action units, and eye gaze contributed 72, 161, 119 features, and 31 features, respectively.

Due to the larger feature space, we applied the Pearson Correlation technique to identify and eliminate redundant features from the summarized features. The feature reduction step was implemented separately for head pose, body pose, and action unit features.  We chose a correlation threshold of 0.75 to discard redundant features in line with the existing literature~\cite{rashid2020predicting,perez2017accurate}. After the feature reduction step, action units, body pose and head pose reduced to 53, 12, and 25, features respectively. 
%were as follows: action units - 92 x 53, body pose - 92 x 12, head pose - 92 x 25.
Table~\ref{tab:non-redundant-features} lists the final set of non-redundant features.
We did not apply feature reduction to sequential features since we intended to use a deep learning model for eye gaze, which can identify essential features independently. Hence, the final size of our combined dataset is 92 x 121.

For label creation, we utilized the total self-reported anxiety score (see section \ref{section_socialanxiety_score_processing}).  Labels were assigned to participants based on their total self-reported post-speech questionnaire scores, where a score above 15 is labeled as `1' (SAD class), and participants with a score equal to or below 15 are labeled as `0' (Non-SAD class). With this paper, we are making our dataset publicly available~\cite{our-video-dataset}. This initiative is essential as there are currently no publicly accessible datasets for video-based SAD analysis. This release aims to support the research community by providing a benchmark for comparing methods and fostering advancements in the field.

\section{Models}
We implemented classical supervised Machine Learning (ML) and supervised Deep Learning (DL) methods for the classification of the participants (SAD versus Non-SAD) using the extracted features. Now, we will be discussing them briefly with respect to the input features.

\noindent{\bf i. Models for summarized features:} 
We employed commonly used machine learning algorithms, including Decision Tree (DT)~\cite{burkov2019hundred}, Gaussian Naive Bayes (GNB)~\cite{bishop2006pattern}, Gradient Boost (GBoost), and Random Forest (RF)~\cite{article_random_forest}. 
%These algorithms were subsequently trained on the designated training dataset and tested on the test set, as detailed in the section \ref{section:dataset}.

\begin{figure}[!h]
    \centering
    \includegraphics[scale=0.34]{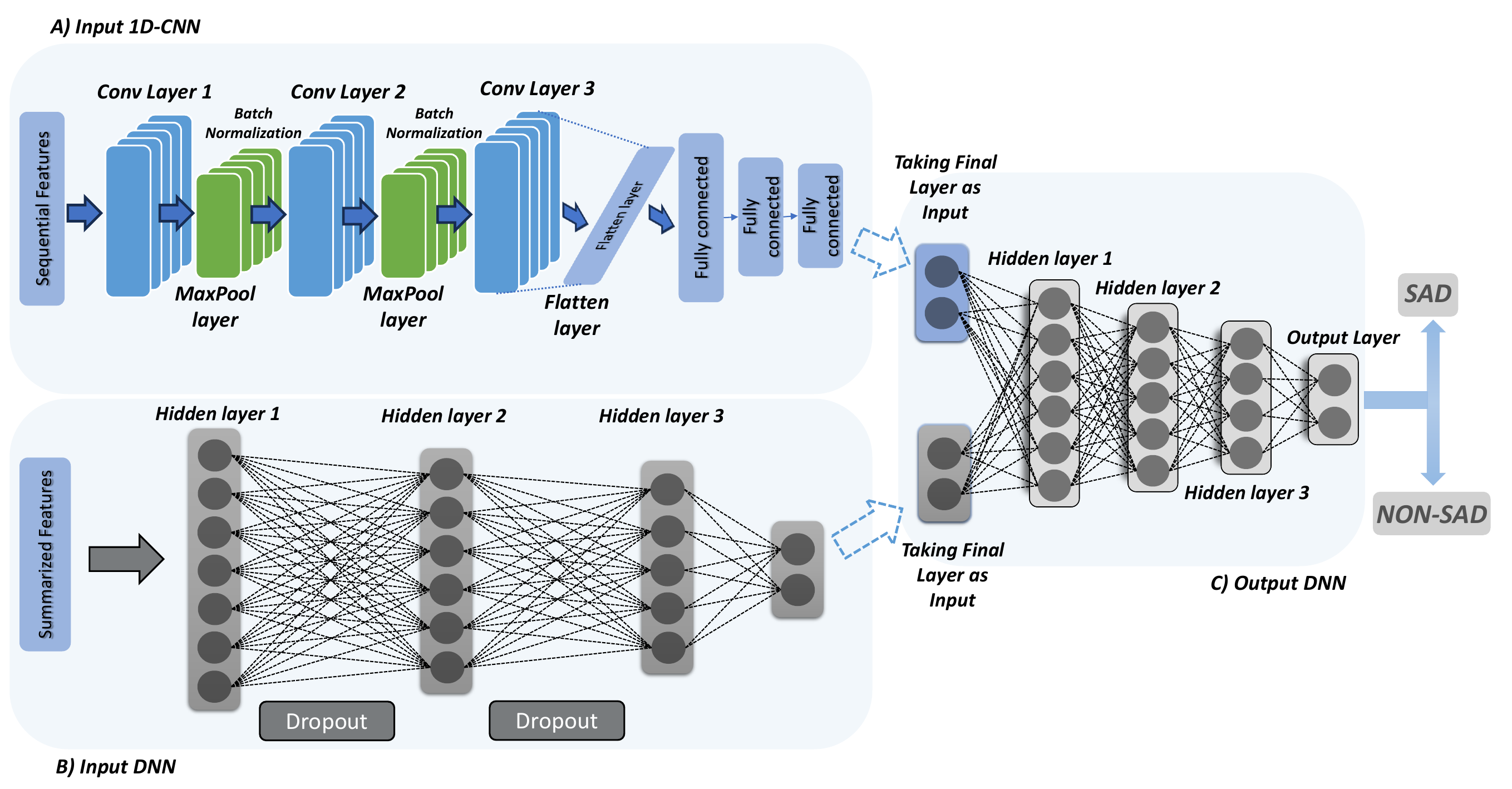}
    \caption{Hybrid 1D CNN-DNN Architecture: Sequential eye gaze features are processed through 1D-CNN, and summarized features are handled by the input DNN. The output from both layers is combined, and inputs to output DNN for the final classification.}
    \label{fig: CNN-DNN architecture}
\end{figure}

Additionally, we developed a custom Deep Neural Network (DNN) model for summarized features, illustrated in Figure \ref{fig: CNN-DNN architecture} (see Input DNN architecture). It takes summarized features derived from head pose (C2), body pose (C3), and action units (C1) as input. The aim was to explore the efficiency of deep learning in classification for summarized features, comparing its evaluation metrics, such as accuracy, with those of the classical ML algorithms.

\noindent{\textit{Parameter setting:}}  The parameters used for DT were \{criterion=`gini', splitter=`best'\}, for RF - \{n\_estimators=50, criterion=`gini'\}, for GBoost - \{n\_estimators=70\}, and for GNB - \{priors=[0.50, 0.50]\}. The other parameters were kept as default. \\

\noindent{\bf ii. Models for sequential features:} 
Classical ML models often struggle with time series data. However, DL models excel in this domain due to their inherent capability to automatically extract meaningful features and learn hierarchical representations. Consequently, we have employed DL methods to classify participants into SAD and non-SAD categories based on sequential eye gaze features. Specifically, we have leveraged, Deep Neural Network(DNN)~\cite{Borisov_2022}, 1D-Convolutional Neural Network (1D-CNN)~\cite{kiranyaz20211d}, along with Long Short Term Memory (LSTM)~\cite{LINDEMANN2021650} for our classification tasks.

\noindent{\textit{Parameter setting:}}  
For all models (DNN, 1D-CNN, and LSTM), we maintained the temporal sequence of participant eye gaze data by passing it sequentially. We divided the data of 92 participants into sequences, each comprising 1978 data points representing the duration of a single participant's video. From the eye gaze data, we extracted 31 distinct features. However, for model training, we specifically utilized the dataset from 69 participants, resulting in an input size of (69, 1978, 31).

We assessed a Deep Neural Network (DNN) model comprising three hidden layers, structured with 64, 32, and 2 neurons in sequential layers. Rectified Linear Unit (ReLU) was the activation function for the first two layers, while the final output layer utilized the sigmoid activation function.

For the 1D-CNN model, we used three convolutional layers. Each layer has several feature maps $N\in \{2,16,32,64,128\}$. Through evaluation, we found optimal feature maps to be 64 and 32. Each convolutional layer has a Rectified Linear Unit (Relu) as the activation function, with the batch normalization and max pooling layer attached after the first and second convolutional layers. The resulting feature vector from the convolution layers was flattened and passed to fully connected layers. Here, we took three fully connected layers with 32, 16, and 2 as the number of neurons with activation function set as  Rectified Linear Unit (ReLU) for the first two layers and sigmoid for the last output layer.

Similarly, for the LSTM model, we used 2 LSTM layers with 64 and 32 neurons each, followed by two fully connected layers. Each layer has 16 and 2 as the number of neurons, with the activation function set as (ReLU) and sigmoid for the first and second fully connected layers. For each of the above-mentioned DL models, we used ``\textbf{Adam}'' as the optimizer with a batch size of 32 employed with ``\textbf{categorical cross entropy}'' as the loss function. \\

\noindent{\bf iii. Model for combined summarized and sequential features:} 
We used an ensemble approach~\cite{10.5555/648054.743935} to combine summarized and sequential features and created a 1D-CNN-DNN architecture shown in Figure~\ref{fig: CNN-DNN architecture}. The 1D-CNN-DNN model comprises of input (i.e., input 1D-CNN \& DNN) and output (i.e., output DNN) sub-modules. Following is the description of each of these sub-modules.  

{\bf Input 1D-CNN:} It processes each participant's eye gaze data sequentially, allowing the model to capture temporal changes across all features. Specifically, we divide all the participant eye gaze data into sequences, where each sequence comprises of 1978 frames, equivalent to the entirety of a single participant's data. We have used 31 distinct features from the eye gaze data. Consequently, the training data is taken as input, resulting in the dimension for the model being structured as (69, 1978, 31), with 69 denoting the total number of participants in the training dataset, 1978 representing the sequence length, and 31 representing the total number of features. This sequential processing ensures that features corresponding to each participant are appropriately fed into the 1D-CNN descriptor, and the subsequent fully connected layers generate precise two-class label prediction, resulting ``(none, 2)'' as the output layer shape, where `none' represents the total samples (can vary), and `2' refers to probabilities scores corresponding to SAD and Non-SAD labels.

{\it Parameter settings:}
For the input 1D-CNN part of the architecture, we evaluated the model with two convolution layers. Where each convolutional layer was configured with a kernel size of 1x3 and a default stride offset of 1. For convolutional layers, we tried the number of feature maps $N\in$ \{16, 32, 64, 128\} and found the optimal results by using 64 and 32 as the number of feature maps. Each layer uses the Rectified Linear Unit (Relu) as the activation function. The batch normalization layer and max-pooling layer, essential components of our architecture, were attached after the first and second convolution layers, where for the max-pooling layer, a kernel size of 1x2 is applied to capture the salient features effectively. The feature vectors obtained from the last convolution layer are then flattened and passed to the fully connected layers. For the fully connected layers, we explored various configurations for the number of nodes $ N\in$ \{2, 16, 32, 64, 128\}, where we got good results from using 32, 16 and 2 as the number of nodes. For the initial two fully connected layers, we used the Rectified Linear Unit (Relu) as the activation function with kernel regularization of type ``L2''~\cite{cortes2012l2}. Notably, the last fully connected layer serves as the output layer. Therefore, the sigmoid activation function is used to facilitate classification. Further, ``\textbf{Adam}'' is used as the optimizer with a batch size of 32, employed with ``\textbf{categorical cross entropy}'' as the loss function. 

{\bf Input DNN:} It takes summarized features derived from head pose, body pose, and action units as input. We have extracted a total of 90 features from the combined head pose (25 features), body pose (12 features), and action units (53 features) dataset. Consequently, the input size for the DNN model will vary based on the features we want to use in the model. For instance, if we take all the summarised features together (AU, HP \& BP - C8), the resulting input for the DNN model will be structured as (69, 90), where 69 signifies the overall participant count, and 90 corresponds to the total number of features. To give final two-class label prediction scores resulting in ``(none, 2)'' where `none' refers to the total number of samples (can vary) and `2' refers to probabilities score corresponding to SAD and Non-SAD label.

{\it Parameter settings:}
For the input DNN part of the architecture, we evaluated the model by exploring different configurations. We varied the number of hidden layers $N \in $ \{2, 3, 4\} and the number of nodes in each layer $N \in $ \{2, 16, 32, 64\}. The optimal results were achieved with three hidden layers, utilizing 64, 32, and 2 nodes in these layers. For the initial two hidden layers, the Rectified Linear Unit (Relu) is used with kernel regularization of type ``L2'', but for the last hidden layer, we used sigmoid as the activation function. After the first and second hidden layers, a dropout layer is used with a ratio of 0.5 to prevent the case of overfitting~\cite{ying2019overview}. Further, ``\textbf{Adam}'' is used as the optimizer with a batch size of 16, employed with ``\textbf{categorical cross entropy}'' as the loss function.

{\bf Output DNN:} Once both input 1D-CNN and DNN have been independently trained, we freeze their weights~\cite{298610}. Furthermore, we extract the last fully connected layers from both models and employ them as inputs to a distinct output DNN. It is worth noting that the output feature vectors from the input 1D-CNN and DNN models exhibit distinct shapes. To facilitate their seamless integration into the subsequent DNN model, we concatenate these two feature vectors into a single feature vector. This combined feature vector is then forwarded to the DNN model, which serves as the final step in our predictive pipeline. Importantly, during the output DNN model training, there is no back-propagation~\cite{118638} occurring from the output DNN to the input 1D-CNN and DNN models. This Final DNN model pipeline is responsible for predicting and classifying participants into either the SAD or Non-SAD categories.

{\it Parameter settings:}
For the output DNN part of the architecture, utilizing the knowledge of the training and testing data composition, we fine-tune the DNN network accordingly. The parameter settings for this segment mirror those used for the standalone input DNN architecture. The only exceptions are the inclusion of a dropout ratio of 0.4 and the introduction of batch normalization between the second and third hidden layers.

% \begin{figure}[h]
%     \centering
%     \includegraphics[scale=0.34]{Figures/architecture/new_diagram.pdf}
%     \caption{Hybrid 1D CNN-DNN Architecture: Sequential eye gaze features are processed through 1D-CNN, and summarized features are handled by the input DNN. The output from both layers is combined, and inputs to output DNN for the final classification.}
%     \label{fig: CNN-DNN architecture}
% \end{figure}

\subsection{Evaluation}
Given the size of our dataset, we utilized k-fold validation \cite{xiong2020evaluating} for training and testing the machine learning models.
K-fold validation involves dividing the dataset into ‘k’ equal parts, where the (k-1) parts are used for training and 1 part is left out for testing in each iteration. The folds are repeated until all parts have been used to train and test the data in each fold, after which the average \textit{accuracy, precision and recall} are reported.

Specifically, we used stratified k-fold \cite{wu2019prediction} approach. It is a variation of the traditional K-fold cross-validation, which also considers the proportion of the class labels present in the dataset. It also divides the data into ‘k’ equal parts, albeit the proportion of the class labels is also considered during the division into folds. For example,  if a dataset has 100 data points, where 60 belong to the ‘0’ class and 40 belong to the ‘1’ class, then in the case of 5 fold stratified k-fold, each fold will contain 20 data points, where 12 points will belong to ‘0’ class and 8 will belong to ‘1’ class.

For our analysis, we divided our dataset into 5 folds for analysis, following which the average accuracy across 5 folds is reported in the paper. For the deep learning part, we utilized the k-fold validation in a similar manner for splitting the data. For training and testing the model, we trained the DNN model on summarised features and the 1D-CNN model on the sequential features on 5 folds, after which we saved the models and weights of the best-performing model. Then, these saved models were used by the combined 1D-CNN-DNN model to make predictions using both summarised and sequential features.

\subsubsection{Metrics}

We have used different evaluation metrics to judge the performance of the mentioned classification methods \cite{hossin2015review}. We used Precision, Recall, F-1 score, and Accuracy metrics. Mathematically,   
\vspace{0.2cm}
% \hl{put metrices in just 2 lines}

\noindent\begin{minipage}{.5\linewidth}
\begin{equation}
  \text{Precision} = \frac{TP}{TP + FP}
\end{equation}
\end{minipage}%
\begin{minipage}{.5\linewidth}
\begin{equation}
  \text{Recall} = \frac{TP}{TP + FN}
\end{equation}
\end{minipage}%

\hspace{0.2cm}

\noindent\begin{minipage}{.5\linewidth}
\begin{equation}
  \text{F1-Score} = \frac{2*Precision*Recall}{Precision + Recall}
 \end{equation}
\end{minipage}%
\begin{minipage}{.5\linewidth}
\begin{equation}
  \text{Accuracy} = \frac{TP+TN}{TP+TN+FP+FN}
\end{equation}
\end{minipage}%
\vspace{0.5 cm}

Here TP, TN, FP, and FN represent true positive, true negative, false positive, and false negative.  In our assessment, a data point qualifies as a `True Positive' (TP) when both the predicted and actual labels are 1, indicating the method's successful classification of a participant as SAD within the SAD class. Conversely, a `True Negative' (TN) occurs when both the predicted and actual labels are 0, signifying the method's accurate classification of an individual as Non-SAD. In addition, a `False Positive' (FP) arises when the actual label is 0, but the predicted label is 1. In our context, this denotes a participant being labeled as `SAD' by the model despite belonging to the `Non-SAD' category. Similarly, a `False Negative' (FN) occurs when a participant with an actual label of 1 (SAD) is classified as belonging to the Non-SAD category (predicted label 0) by the model.

\section{Results}
This section presents the SAD versus non-SAD classification results from different classification models. For brevity, we use abbreviations (i.e., $C_{i}$) for features and their combinations, as shown in Table~\ref{tab: comb} in the rest of the paper. Further, the table also shows the classification models applied to the respective features. Following is a brief description of the classification results obtained on summarized, sequential, and the combination of summarized and sequential features. 

\begin{table}[h]
 \tiny
\caption{Different combination of features}
\label{tab: comb}
\begin{tabular}{@{}lll@{}c}
\toprule
%                                  & \multicolumn{2}{c}{\textbf{Features}}                     \\ 
\multicolumn{1}{l|}{\textbf{Abbr.}}  & \multicolumn{2}{l}{\bf Features} &                                   Classification Models\\
\midrule
\multicolumn{1}{l|}{\textbf{C1}}  & \multicolumn{2}{l}{Action Units (AUs)} &                      \\
\multicolumn{1}{l|}{\textbf{C2}}  & \multicolumn{2}{l}{Head Pose (HP)} &                             ML and DNN\\
\multicolumn{1}{l|}{\textbf{C3}}  & \multicolumn{2}{l}{Body Pose (BP)} &                             \\ \midrule
\multicolumn{1}{l|}{\textbf{C4}}  & \multicolumn{2}{l}{Eye Gaze (EG)} &                              1D-CNN, LSTM \& DNN\\ \midrule

\multicolumn{1}{l|}{\textbf{C5}}  & \multicolumn{2}{l}{AUs + Head Pose} &                       \\
\multicolumn{1}{l|}{\textbf{C6}}  & \multicolumn{2}{l}{AUs + Body Pose} &                       ML and DNN\\
\multicolumn{1}{l|}{\textbf{C7}}  & \multicolumn{2}{l}{Head Pose + Body Pose} &                 \\
\multicolumn{1}{l|}{\textbf{C8}}  & \multicolumn{2}{l}{AUs + Head Pose + Body Pose} &                     \\ \midrule
\multicolumn{1}{l|}{\textbf{C9}} & \multicolumn{2}{l}{Head Pose + Eye Gaze} &                  \\
\multicolumn{1}{l|}{\textbf{C10}} & \multicolumn{2}{l}{Eye Gaze + Body Pose} &                  \\
\multicolumn{1}{l|}{\textbf{C11}} & \multicolumn{2}{l}{AUs + Eye Gaze} &           \\
\multicolumn{1}{l|}{\textbf{C12}} & \multicolumn{2}{l}{AUs + Eye Gaze + Head Pose} &            1D-CNN-DNN\\
\multicolumn{1}{l|}{\textbf{C13}} & \multicolumn{2}{l}{AUs + Eye Gaze +Body Pose} &             \\
\multicolumn{1}{l|}{\textbf{C14}} & \multicolumn{2}{l}{Eye Gaze + Body Pose + Head Pose} &      \\
\multicolumn{1}{l|}{\textbf{C15}} & \multicolumn{2}{l}{AUs + Body Pose +Head Pose + Eye Gaze} & \\ \bottomrule
\end{tabular}

\end{table}
 
%This section presents and interprets our thorough analysis of diverse features extracted from the videos. To enhance clarity and organization, we have structured this section into three parts, each focusing on distinct features used to train the models. Firstly, we present the performance of summarized features, i.e., C1, C2, C3, C5, C6, C7, C8 using both ML and DL techniques. Next, we delve into the sequential feature, and we further present the performance of the proposed DL architecture trained and tested on combined summarized and sequential features.

\begin{figure}
  \centering
  
  \begin{subfigure}{0.3\textwidth}
    \centering
    \includegraphics[scale=0.25]{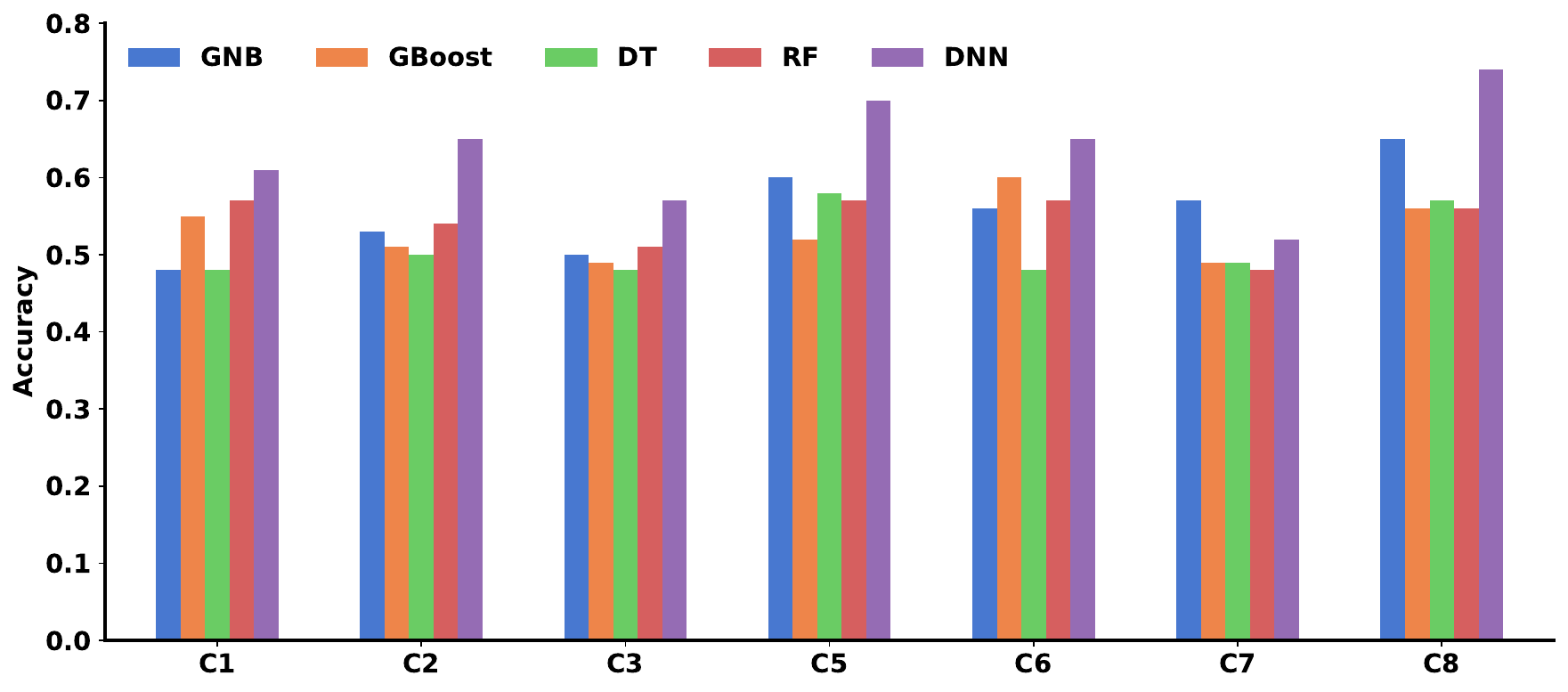}
    \caption{}
    \label{fig:bar-plot-accuracy-summarized}
  \end{subfigure}
  \hfill
  \begin{subfigure}{0.2\textwidth}
    \centering
    \includegraphics[scale=0.25]{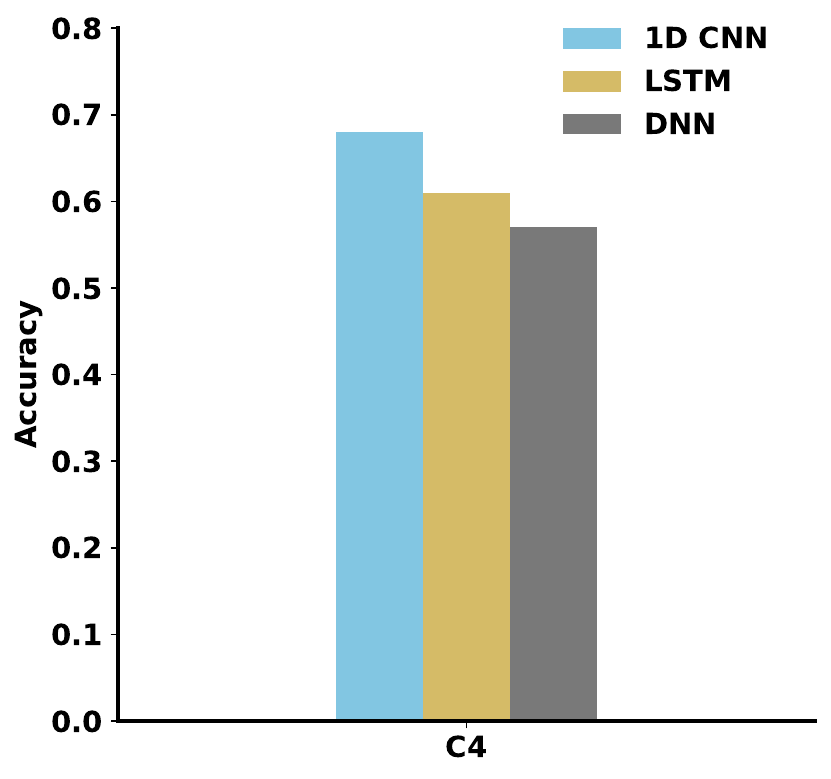}
    \caption{}
    \label{fig:bar-plot-eye-gaze}
  \end{subfigure}
  \hspace{0.001\textwidth}
  \begin{subfigure}{0.3\textwidth}
    \centering
    \includegraphics[scale=0.25]{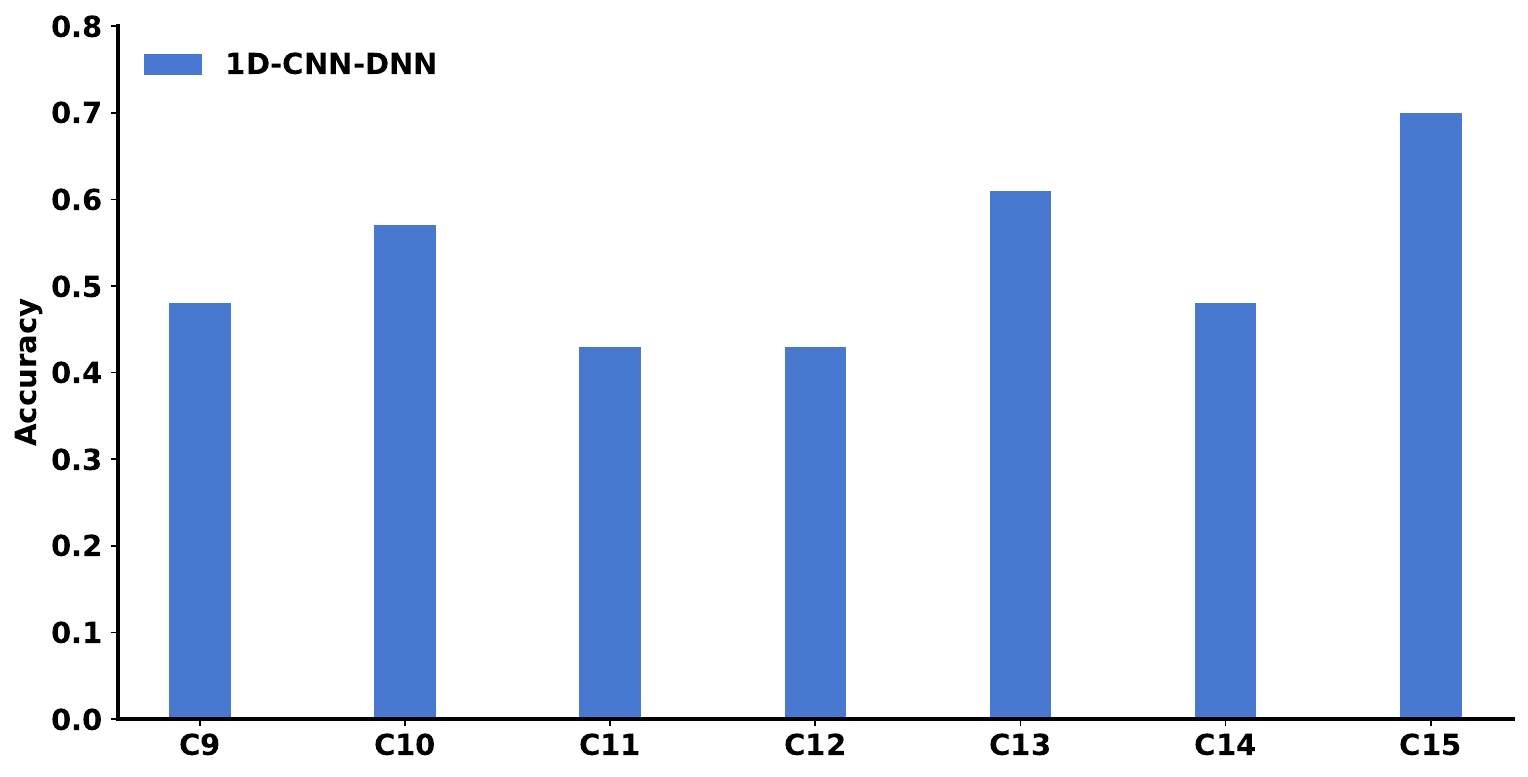}
    \caption{}
    \label{fig:bar-plot-accuracy-combined}
  \end{subfigure}

  \caption{Accuracies (\%) on (a) Summarized features with Gaussian Naive Bayes (GNB), Decision Tree (DT), Random Forest (RF), and Gradient Boost (GBoost),  Deep Neural Network (DNN), (b) Eye gaze features with 1D-CNN, LSTM, and DNN, (c) Combined summarized and sequential features with 1D-CNN-DNN. Abbreviations (i.e., $C_i$) for feature combinations are provided in Table~\ref{tab: comb}.}
  \label{fig:accuracy-plots}
\end{figure}

\noindent{\bf Summarized features:} Figure~\ref{fig:bar-plot-accuracy-summarized} presents the classification accuracies obtained with Gaussian Naive Bayes (GNB), Decision Tree (DT), Random Forest (RF), and Gradient Boost (GBoost), and Deep Neural Network (DNN) models on summarized features, C1, C2, C3, C5, C6, C7, and C8.

The highest and lowest accuracy achieved by DNN were 74\% (on C8) and 52\% (on C7), respectively. In summary, the proposed DNN outperformed classical ML models on all summarized feature combinations except for C7. This suggests that the deep neural networks was more adept at learning patterns compared to classical ML methods. \\

\begin{comment}
    \begin{figure}
  \begin{subfigure}{0.5\textwidth}
    \centering
    \includegraphics[width=\linewidth]{new_ACM_plots/C8_loss_plot_bold.pdf}
    \caption{Loss Plot for C8}
    \label{fig:figure1}
  \end{subfigure}%
  \begin{subfigure}{0.5\textwidth}
    \centering
    \includegraphics[width=\linewidth]{new_ACM_plots/C15_loss_plot_bold.pdf}
    \caption{Loss Plot for C15}
    \label{fig:figure2}
  \end{subfigure}
  \caption{Loss plots for features C8 and C15}
  \label{fig:combined}
\end{figure}
\end{comment}

\noindent{\bf Sequential features:}
Figure~\ref{fig:bar-plot-eye-gaze} illustrates the accuracies achieved with 1D-CNN, LSTM, and DNN exclusively using Eye gaze (C4) features. The 1D-CNN model, yielded the highest accuracy of 68\% along with superior precision, recall, and F1-score than the LSTM and DNN as shown in the Table\ref{tab:c9_c15}.

\noindent{\bf Combined summarized and sequential features:} Figure~\ref{fig:bar-plot-accuracy-combined} and Table~\ref{tab:c9_c15} present the accuracy of the 1D-CNN-DNN model across various combinations (C9, C10, C11, C12, C13, C14, and C15) of summarized and sequential features. In this setup, the 1D-CNN architecture was employed to train the sequential features, while the DNN was utilized to train the summarized features. These features were subsequently fused and fed into another DNN model. The results indicate that the model achieved its highest classification accuracy of 70\% on the C15 combination, with a precision and recall of 0.69 and 0.65, respectively.

\begin{table} []
\caption{F1-score, precision, recall, and accuracy of (a) best ML classifiers and DNN for individual and combined \textit{summarized} feature combinations (i.e., C1, C2, C3, C5, C6, C7, C8) (b) all feature combinations involving eye-gaze \textit{sequential} feature (i.e., C4, C9, C10, C11, C12, C13, C14, C15). Note that  (a) reports results obtained with the best ML classifier for the particular feature combination.}

%\caption{F1-score, precision, recall, and accuracy of (a) best ML classifiers for individual and combined \textit{summarized} feature combinations (i.e., C1, C2, C3, C5, C6, C7, C8) (b) all feature combinations involving eye-gaze \textit{sequential} feature (i.e., C4, C9, C10, C11, C12, C13, C14, C15). Note that  (a) shows results obtained with the best classifier for the particular feature combination.}

\centering
\tiny
\subcaptionbox{\label{tab:c1_c8}}{
\begin{tabular}{lccccccccccccccc}
    \hline
    \multirow{2}{*}{\textbf{}} & \textbf{Features} & \multicolumn{2}{c}{\textbf{C1}} & \multicolumn{2}{c}{\textbf{C2}} & \multicolumn{2}{c}{\textbf{C3}} & \multicolumn{2}{c}{\textbf{C5}} & \multicolumn{2}{c}{\textbf{C6}} & \multicolumn{2}{c}{\textbf{C7}} & \multicolumn{2}{c}{\textbf{C8}} \\
    \cmidrule{2-2} \cmidrule(lr){3-4} \cmidrule(lr){5-6} \cmidrule(lr){7-8} \cmidrule(lr){9-10} \cmidrule(lr){11-12} \cmidrule(lr){13-14} \cmidrule(lr){15-16}
    % \cline{2-16}
     & \textbf{Model}  & \textbf{RF} & \textbf{DNN} & \textbf{GNB} & \textbf{DNN} & \textbf{DT} & \textbf{DNN} & \textbf{DT} & \textbf{DNN} & \textbf{GBoost} & \textbf{DNN} & \textbf{GNB} & \textbf{DNN} & \textbf{GNB} & \textbf{DNN} \\ 
    \hline
    %\textbf{Model} & & & & & & & & & & & & & & & \\
    \textbf{F1-Score} & & 0.58 & 0.60 & 0.56 & 0.63 & 0.51 & 0.56 & 0.59 & 0.67 & 0.55 & 0.63 & 0.56 & 0.51 & 0.63 & 0.71 \\
    \textbf{Precision} & & 0.58 & 0.60 & 0.61 & 0.63 & 0.52 & 0.61 & 0.60 & 0.68 & 0.56 & 0.63 & 0.58 & 0.51 & 0.67 & 0.73 \\
    \textbf{Recall} & & 0.57 & 0.60 & 0.58 & 0.63 & 0.52 & 0.60 & 0.60 & 0.68 & 0.58 & 0.63 & 0.57 & 0.51 & 0.64 & 0.71\\
    \textbf{Accuracy} & & 0.57 & 0.61 & 0.58 & 0.65 & 0.53 & 0.57 & 0.60 & 0.70 & 0.58 & 0.65 & 0.57 & 0.52 & 0.64 & 0.74 \\
    \hline
\end{tabular}
}
\hspace{1 cm}
\subcaptionbox{\label{tab:c9_c15}}{
\begin{tabular}{cclccccccl} \toprule
         \textbf{Features} 
 &\textbf{C4} 
 &&  \textbf{C9} &  \textbf{C10} &  \textbf{C11} &  \textbf{C12} &  \textbf{C13} & \textbf{C14} &\textbf{C15} 
\\ \midrule
         \textbf{Model}  & \textbf{1D-CNN } 
 && \multicolumn{7}{c}{\bf 1D-CNN-DNN} \\ \cmidrule{2-2} \cmidrule(r){4-10}
 
         \textbf{F1-score}        
 &0.77
 &&  0.45 &  0.54    &  0.36   &  0.43   &  0.60  &  0.42    &0.65
\\ 
         \textbf{Precision}       
 &0.63
 &&  0.59 &  0.74    &  0.70   &  0.35   &  0.75  &  0.71    &0.69
\\ 
         \textbf{Recall}          
 &0.63
 &&  0.55 &  0.64  &  0.54   &  0.38   &  0.68  &  0.57    &0.65
\\ 
         \textbf{Accuracy}        
 &0.68 &&  0.48 &  0.57  &  0.43   &  0.43   &  0.61  &  0.48    &0.70\\ \bottomrule
    \end{tabular}

}
%\addlinespace[0.25cm]
\label{tab:k_fold_results}

\end{table}

\section{Discussion}  
In this study, we predicted social anxiety by leveraging different bodily features extracted from participants' videos. Initially, we extracted body pose (BP), head pose (HP), action units (AUs), and eye gaze (EG) features from each participant's video collected during speech activity. Subsequently, we employed correlation techniques to identify redundant features. Finally, we tested whether selected non-redundant features can accurately detect whether a participant is suffering from social anxiety or not.

\subsection{Impact of feature combinations on classification accuracy}
The correlation technique played a crucial role in feature reduction. Initially, the number of features corresponding to 92 participants were 72, 161, and 119 for the HP, BP, and AUs, respectively. The correlation techniques reduced the features to 25, 12, and  53 for HP, BP, and AUs, respectively. Feature reduction is essential as unwanted features affect the model accuracy and increase computation time. Further, we experimented with both individual sets of features (i.e., standalone HP, BP, AUs, and EG) and combinations of features, such as the fusion of HP and BP, fusion of HP, BP, and AUs, etc. (see table \ref{tab: comb}). The objective was to assess the significance of HP, BP, AUs, and EG features at individual and fusion levels in detecting SAD. This approach is guided by the diverse literature where some studies focused solely on body pose~\cite{horigome2020evaluating}, action units~\cite{gavrilescu2019predicting, zhang2020video}, or a combination of AUs and BP \cite{liu2022measuring}. Now, we will discuss the standalone and fusion-level feature findings in detail.

\subsubsection{Standalone features}
We applied both classical ML and deep learning techniques to standalone features. Classical ML models and DNN were applied to summarized features AU (i.e., C1), HP (C2), and BP (C3), while the 1D-CNN was applied to Eye gaze (C4). Our findings revealed that the DNN trained on standalone summarized features outperformed classical ML models. The DNN trained on HP (C2) achieved the highest accuracy of 65\%, surpassing AUs (61\%) and BP (57\%). Furthermore, this suggests that the DNN model was able to discern discriminating patterns in head movements between SAD and non-SAD participants more appropriately than AUs and BP. Similarly, the 1D-CNN model trained on eye gaze (C4) achieved the highest accuracy of 68\% in discriminating between SAD and non-SAD participants.

\subsubsection{Fused feature combinations}
The fusion of features (C5 to C15) displayed a varied impact on classification accuracy, with some combinations showing an increase and others a decrease. The highest accuracy of 74\% was achieved with C8 (AUs + HP + BP) using DNN. Regarding summarized features, we observed increased accuracy with the fusion of two or more summarized features, except for the combination of HP and BP. Table \ref{tab:fusion_duo_trio_quad} illustrates the change in accuracy, mostly showing an increase in the case of summarized features. For instance, AUs, HP, and BP alone yielded accuracies of 61\%, 65\%, and 57\%, respectively. However, when AUs and HP were fused, the accuracy increased to 70\%. Similar improvements were observed for AUs and BP. Furthermore, when AUs, HP, and BP were fused together, the highest accuracy of 74\% was achieved.

Additionally, we noted that the inclusion of EG led to a drop in accuracy, suggesting that the sequential feature, i.e., EG alone, is effective but not when combined with summarized features. In summary, AUs, BP, and HP combined were identified as the most prominent feature fusions in discriminating SAD from non-SAD participants.

\begin{comment}
    The fusion of features (C5 to C15) demonstrated a varied impact on classification accuracy, with some combinations showing an increase and others a decrease. The highest accuracy of 74\% was achieved with C8 (AUs + Head Pose + Body Pose) using DNN. 

Similarly, an accuracy of 83\% was achieved with C13 (AUs + Eye Gaze + Body Pose) using 1D-CNN-DNN. Notably, any combination of two features (C6, C7, C9, C10, C11), except for C5, performed less effectively than the individual features. On the other hand, combinations of three features (C8, C12, C13), except for C14, outperformed individual features. Furthermore, on analyzing the change in accuracy from C1 to C15, it was observed that in all cases, Body Pose (BP) combined with other features increased accuracy. Similar patterns were observed for Eye Gaze (EG), except for C12. Moreover, in almost all cases, Action Units (AU) combined with other features increased accuracy, except for two cases (C6 and C11). In summary, AU, BP, and EG combined were identified as the most prominent feature fusions in discriminating SAD from non-SAD participants.
\end{comment}

\begin{table}
    \centering
        \caption{ SAD versus non-SAD classification accuracies and their increase/decrease on transitioning from standalone to the fusion of two, three, and four feature combinations. $\uparrow$ and $\downarrow$ represent increase and decrease in accuracy, respectively.}
    \label{tab:fusion_duo_trio_quad}
    \begin{tabular}{cc|cc|cc|cc} 
    \toprule
         Standalone&  Acc (\%)&  Fusion (Two)&  Acc ($\uparrow$ / $\downarrow$)&  Fusion (Three)&  Acc ($\uparrow$ / $\downarrow$)&  Fusion (Four)& Acc ($\uparrow$ / $\downarrow$)
\\ \toprule
         AU&  61&  AU + HP&  70 (\textbf{$\uparrow$})&  AU + HP + BP&  74 ($\uparrow$)&  AU + HP + BP +EG& 70 (\textbf{$\downarrow$})
\\ 
         &  &  AU + BP&  65 (\textbf{$\uparrow$})&  AU + HP + EG&  43 (\textbf{$\downarrow$})&  & 
\\ 
         &  &  AU + EG&  43 (\textbf{$\downarrow$})&  AU + BP + EG&  61 (\textbf{$\downarrow$})&  & 

\\ \midrule
         HP&  65&  HP + BP&  52 (\textbf{$\downarrow$})&  HP + BP + EG&  48 (\textbf{$\downarrow$})&  & 
\\ 
         &  &  HP + EG&  48 (\textbf{$\downarrow$})&  &  &  & 

\\  \midrule
         BP&  57&  BP + EG&  57 (\textbf{$\downarrow$})&  &  &  & 
\\ \midrule
 EG& 68
& & & & & &\\ \bottomrule
    \end{tabular}

\end{table}

\subsection{Machine learning versus deep learning methods}
%During feature extraction, we calculated both summarized and sequential features. Summarized features were extracted from action units, body pose, and head pose, while sequential features were extracted from eye gaze. This differentiation was necessary because action units, body pose, and head poses do not change frequently, whereas eye gaze variations change with time (i.e., frames).

For summarized features, we used classical machine learning algorithms (Decision Tree, Gaussian Naive Bayes, Gradient Boost, and Random Forest) and DNN for classification. Conversely, for sequential features, we utilized deep learning techniques such as LSTM (Long Short-Term Memory), 1D-CNN (1D-Convolutional Neural Network), and DNN (Deep Neural Network). The decision to use deep learning techniques aligns with the recent trend in the research community~\cite{8344107,liu2022measuring}.

Furthermore, we designed a 1D-CNN-DNN architecture for the combination of summarized and sequential features. This architecture includes a 1D-CNN network for eye gaze features and a DNN network for handling summarized features, facilitating the fusion of summarized and sequential features. The results indicate that deep learning techniques outperform classical machine learning algorithms, implying that DL techniques are more adept at learning discriminating patterns compared to classical ML algorithms. 

Interestingly, in the case of DL techniques, the 1D-CNN-DNN performed less effectively than the basic DNN. The highest accuracy of 74\% was achieved when trained on DNN alone, whereas the highest accuracy achieved with 1D-CNN-DNN was 70\%. Moreover, upon inspecting the confusion matrix, we observed that although the DNN (trained on C8) misclassified 6 out of 10 non-SAD participants as SAD, it correctly classified 8 out of 9 SAD participants. Conversely, in the case of 1D-CNN-DNN (trained on C15), it misclassified most SAD participants as non-SAD while correctly classifying non-SAD participants. \textit{Considering the low Type II error in the case of DNN, particularly when trained on C8, we conclude that DNN performed the best}.

\begin{comment}
    By using the 1D-CNN-DNN architecture, we achieved the highest accuracy of 83\% in C13 (action unit, body pose, and eye gaze), which was only 1\% less than the highest accuracy observed with Gaussian Naive Bayes in C8 features. In summary, both machine learning and deep learning performed equally well on summarized (C3) and sequential features (C4), as well as on the fusion of features (C13).
\end{comment}

\subsection{Eye gaze}
Eye gaze behavior in individuals with SAD has been extensively examined in previous studies, revealing that socially anxious individuals tend to avoid direct eye contact~\cite{kim2018aversive, weeks2019fear, chen2017gaze, schneier2011fear, schulze2013gaze}. Furthermore, the literature highlights that differences in eye gaze properties between socially anxious and control groups can serve as a discriminating factor.

It is noteworthy that prior studies have utilized high-end eye-tracking devices, such as Kim et al.~\cite{kim2018aversive} used Arrington Research ViewPoint EyeTracker Systems, and Weeks et al.~\cite{weeks2019fear} used D6 high-speed eye tracking systems. In contrast, our approach utilized low-end smartphone camera and  OpenFace library to detect the X, Y, and Z coordinates of the left and right eye, from which various eye gaze features were derived, as discussed in section~\ref{Sequential Features}.

Our findings indicate that eye gaze features alone outperform the standalone AU, BP, and HP features. We achieved an accuracy of 68\% with 1D-CNN using eye gaze features alone. However, combining eye gaze with summarized features yielded poor performance, resulting in a drop in accuracy in classifying SAD versus non-SAD participants. In summary, the eye gaze behavior played a crucial role in distinguishing between SAD and non-SAD participants.

\subsection{Anxiety versus Depression} 
 
 Existing research on behavioral features has focused on stress and depression detection \cite{sun2022estimating,giannakakis2017stress, horigome2020evaluating, gavrilescu2019predicting}. In contrast, our work is the first to use behavioral features to detect social anxiety disorder. Social anxiety and depression are distinct mental health disorders with unique manifestations that impact an individual's thoughts, emotions, and behaviors. Anxiety is typically characterized by excessive worry and fear of negative evaluation, while depression entails persistent sadness, emptiness, and a disinterest in previously pleasurable activities. Physical symptoms associated with each disorder also differ. Anxiety may manifest as restlessness, muscle tension, increased heart rate, sweating, and trembling, whereas depression may result in appetite changes, disrupted sleep patterns (insomnia or hypersomnia), and bodily discomfort.

The nature of anxiety, marked by anticipatory worry and momentary events, contrasts with the more enduring and persistent nature of depression, which can last for extended periods. Given these distinctions between anxiety and depression, comparing the classification accuracy results of our study with prior work on depression detection using behavioral data (action unit, body pose, head pose, eye gaze) proves challenging. Furthermore, while inducing anxiety in a controlled lab environment is facilitated by introducing external stressors, inducing depression in such a setting is challenging.

\subsection{Trait versus state social anxiety}

In our study, we gathered data using two different self-reported measures: the Social Phobia Inventory (SPIN) and a custom questionnaire (see Figure \ref{fig:StudyFlow}). Participants first completed the SPIN, followed by our questionnaire after the speech activity. The SPIN provided insight into a participant's general social anxiety tendencies, while our questionnaire captured their anxiety levels specifically during the speech activity.

Although we could have used the SPIN to classify participants into groups with and without Social Anxiety Disorder (SAD), this method wouldn't have been accurate. The SPIN covers broader aspects of participants' real-life experiences, whereas our questionnaire delves into their feelings and experiences during the specific activity. As a result, we relied on scores from our questionnaire to determine whether participants fell into the SAD or non-SAD categories.

We further analyzed the relationship between SPIN scores and our questionnaire scores, unveiling a strong positive correlation of 0.51, significant at the 0.01 level. This result indicates that our questionnaire adeptly captured participants' social anxiety, highlighting a particular emphasis on their current, momentary social anxiety levels.

\subsection{Implications}
The key takeaways from the findings are listed below:
\begin{itemize}
\item Eye gaze demonstrated highest SAD versus non-SAD classification performance among the individual features. This finding is in sync with the existing literature~\cite{kim2018aversive, weeks2019fear, chen2017gaze, schneier2011fear, schulze2013gaze}.
\item Other key features for identifying social anxiety in participants encompass a combination of action units, body pose, and head pose. Notably, eye gaze, when combined with these individual features decreased classification accuracy.
\item Our analysis revealed deep learning techniques as superior to classical machine learning for classifying SAD versus non-SAD.   
\item Our study introduces a non-invasive approach applicable to real-world scenarios for detecting participant anxiety.
\end{itemize}

\subsection{Limitations and Future Work}

%We used the SPIN questionnaire to group the study participants into SAD and Non-SAD categories. It is also worth noting that our study primarily involved students, and there is a possibility that the SPIN may not have been able to accurately capture the distinctions between SAD and Non-SAD groups in this specific context. Also, 

In our study, it is important to note that we induced anxiety in participants within a controlled laboratory setting using a speech activity. However, this could restrict the generalization of our research findings to real-world situations. The primary limitation stems from the fact that anxiety experienced in everyday life can be considerably more intense and unpredictable compared to the controlled environment of a laboratory. Real-world anxiety can be influenced by various factors such as personal relationships, workplace stress, or financial difficulties, which may not fully replicate in a laboratory setting.

In our study, we have highlighted how bodily indicators like head pose, body pose, action units, and eye gaze can aid in identifying Social Anxiety Disorder (SAD) in individuals. To expand our research, we could include additional features such as blink rate, pupil dilation size, and other anxiety-related biomarkers. This might significantly enhance the accuracy of detecting SAD in individuals. Moreover, this approach could be broadened to identify other mental health conditions like depression and stress. Studies have indicated a strong correlation among stress, anxiety, and depression, allowing for a potentially comprehensive extension of our methodology to encompass these related disorders.

\section{Conclusion}

Our research investigated the efficacy of utilizing various bodily features — head pose, body pose, action units (facial features), and eye gaze — from participants' video data to predict Social Anxiety Disorder (SAD). Our findings reveal that specific combinations of these bodily features outperform other features in detecting SAD. Additionally, deep learning techniques demonstrate better performance when classifying participants into SAD and non-SAD categories. Our analysis suggests that employing non-invasive and privacy-preserving methods for SAD detection, utilizing video data, could significantly enhance the scalability of SAD detection methods.

% \begin{figure}
%   \begin{subfigure}{0.3\textwidth}
%     \centering
%     \includegraphics[scale=0.32]{new_ACM_plots/ACM_results_plot_A_part_new.pdf}
%     \caption{}
%     \label{fig:bar-plot-accuracy-summarized}
%   \end{subfigure}
%   \hfill
% \begin{subfigure}{0.2\textwidth}
%     \centering
%     \includegraphics[scale = 0.2]{new_ACM_plots/ACM_results_plot_B_part_new.pdf}
%     \caption{}
%     \label{fig:bar-plot-eye-gaze}
% \end{subfigure}
%   \hfill
%   \begin{subfigure}{0.40\textwidth}
%     \centering
%     \includegraphics[scale=0.30]{new_ACM_plots/ACM_results_plot_C_part_new.pdf}
%     \caption{}
%     \label{fig:bar-plot-accuracy-combined}
%   \end{subfigure}
%   \caption{Accuracy on (a) Summarized features, (b) Eye gaze features, (c) Summarized and sequential features.}
%   \label{fig:accuracy-plots}
% \end{figure}

\bibliographystyle{unsrt} % Choose your desired bibliography style
\bibliography{references}  

\begin{comment}
    \subsection{Todo}
 1- Is there a way to show the relative importance of features in the hybrid architecture?
2- explore deep learning
 3- Redraw the loss curves. Current ones show overfitting.
 4- Understand the false positives and false negatives.

\end{comment}

\end{document}

\maketitle
\def\@copyrightspace{\relax}

\section{Additional results}
This section presents the detailed results of all computed feature combinations (C1  to C15) in terms of precision, recall, F1-score, and accuracy. 
Table \ref{tab: app_ind} showcases the individual feature performance, encompassing Action units, head pose, body pose, and eye gaze. Subsequently, Table \ref{tab: app_comb} reports the performance of various summarized feature combinations only, offering insights into their combined impact. Finally, Table \ref{tab: app_ey} illustrates the performance of both summarized and eye gaze features, showing the impact of combining summarized and sequential features together.

\section{\textbf{NEW TABLE}}

\vspace{0.4cm}
\begin{table}[h]
\caption{Precision, recall, F1-score and accuracy of the different classifiers on the individual features only.}
\label{tab: app_ind}
\begin{tabular}{@{}lllcccc@{}}
\toprule
\textbf{Acronym} & \textbf{Features} & \textbf{Methods} & \textbf{Precision} & \textbf{Recall} & \textbf{F1-score} & \textbf{Accuracy} \\ \midrule
\multirow{4}{*}{\textbf{C1}} & \multirow{4}{*}{\textbf{Action Units (AUs)}} & Decision Tree & 0.48 & 0.48 & 0.47 & 0.48 \\
 & & Random Forest & 0.58 & 0.57 & 0.58 & 0.58 \\
 & & Gradient Boost & 0.54 & 0.53 & 0.52 & 0.54 \\
 & & Gaussian Naive Bayes & 0.55 & 0.55 & 0.50 & 0.55 \\ \midrule
\multirow{4}{*}{\textbf{C2}} & \multirow{4}{*}{\textbf{Head Pose}} & Decision Tree & 0.51 & 0.51 & 0.50 & 0.50 \\
 & & Random Forest & 0.52 & 0.52 & 0.51 & 0.52 \\
 & & Gradient Boost & 0.54 & 0.53 & 0.52 & 0.53 \\
 & & Gaussian Naive Bayes & 0.61 & 0.58 & 0.56 & 0.58 \\ \midrule
\multirow{4}{*}{\textbf{C3}} & \multirow{4}{*}{\textbf{Body Pose}} & Decision Tree & 0.52 & 0.52 & 0.51 & 0.53 \\
 & & Random Forest & 0.50 & 0.49 & 0.48 & 0.49 \\
 & & Gradient Boost & 0.52 & 0.51 & 0.50 & 0.51 \\
 & & Gaussian Naive Bayes & 0.50 & 0.49 & 0.46 & 0.49 \\ \midrule
\multirow{3}{*}{\textbf{C4}} & \multirow{3}{*}{\textbf{Eye Gaze}} & DNN & 0.59 & 0.57 & 0.55 & 0.57 \\
 & & LSTM & 0.55 & 0.51 & 0.44 & 0.61 \\
 & & 1D-CNN & 0.63 & 0.63 & 0.77 & 0.68 \\ \bottomrule
\end{tabular}
\end{table}

\vspace{0.4cm}
\begin{table}[h]
\caption{Precision, recall, F1-score, and accuracy of different classifiers on various combinations of summarized features.}
\begin{tabular}{@{}lllcccc@{}}
\toprule
\textbf{Acronym}              & \textbf{Features}                                                                                                    & \textbf{Methods}    & \multicolumn{1}{l}{\textbf{Precision}} & \multicolumn{1}{l}{\textbf{Recall}} & \multicolumn{1}{l}{\textbf{F1-socre}} & \multicolumn{1}{l}{\textbf{Accuracy}} \\ \midrule
\multirow{4}{*}{\textbf{C5}}  & \multirow{4}{*}{\textbf{\begin{tabular}[c]{@{}l@{}}Action Units \\ (AUs) +\\ Head Pose\end{tabular}}}                & Decision Tree       & 0.60& 0.60& 0.59& 0.60\\
                              &                                                                                                                      & Random Forest                 & 0.53& 0.53& 0.51& 0.53\\
                              &                                                                                                                      & Gradient Boost                 & 0.53                                   & 0.52                                & 0.52                                  & 0.57                                  \\
                              &                                                                                                                      & Gaussian Naive Bayes & 0.49                                   & 0.49                                & 0.49                                  & 0.50                                  \\ \midrule
\multirow{4}{*}{\textbf{C6}}  & \multirow{4}{*}{\textbf{\begin{tabular}[c]{@{}l@{}}Action Units\\  (AUs) +\\ Body Pose\end{tabular}}}                & Decision Tree       & 0.46& 0.48& 0.45& 0.48\\
                              &                                                                                                                      & Random Forest                 & 0.54& 0.55& 0.53& 0.55\\
                              &                                                                                                                      & Gradient Boost                 & 0.56& 0.58& 0.55& 0.58\\
                              &                                                                                                                      & Gaussian Naive Bayes & 0.60& 0.55& 0.51& 0.55\\ \midrule
\multirow{4}{*}{\textbf{C7}}  & \multirow{4}{*}{\textbf{\begin{tabular}[c]{@{}l@{}}Head Pose + \\ Body Pose\end{tabular}}}                           & Decision Tree       & 0.49& 0.49& 0.48& 0.49\\
                              &                                                                                                                      & Random Forest                 & 0.49& 0.49& 0.48& 0.49\\
                              &                                                                                                                      & Gradient Boost                 & 0.52& 0.52& 0.51& 0.52\\
                              &                                                                                                                      & Gaussian Naive Bayes & 0.58& 0.57& 0.56& 0.57\\ \midrule
\multirow{4}{*}{\textbf{C8}} & \multirow{4}{*}{\textbf{\begin{tabular}[c]{@{}l@{}}Action Units \\ (AUs) +\\ Body Pose + \\ Head Pose\end{tabular}}} & Decision Tree       & 0.60                                   & 0.60                                & 0.60                                  & 0.60\\
                              &                                                                                                                      & Random Forest                 & 0.65                                   & 0.65                                & 0.65                                  & 0.65                                  \\
                              &                                                                                                                      & Gradient Boost                 & 0.72                                   & 0.63                                & 0.62                                  & 0.70                                  \\
                              &                                                                                                                      & Gaussian Naive Bayes & 0.67& 0.64& 0.63& 0.64\\ \bottomrule
\end{tabular}
\label{tab: app_comb}
\end{table}

\vspace{0.5cm}

\begin{table}[]
\centering      
\caption{Precision, recall, F1-score, and accuracy of the best-performing classifiers (1D-CNN-DNN) for different combination of summarized and sequential features.}
\label{tab: app_ey}

\begin{tabular}{@{}llcccc@{}}
\toprule
\textbf{Acronym} & \textbf{Features}                                                                                         & \multicolumn{1}{l}{\textbf{Precision}} & \multicolumn{1}{l}{\textbf{Recall}} & \multicolumn{1}{l}{\textbf{F1-score}} & \multicolumn{1}{l}{\textbf{Accuracy}} \\ \midrule
\textbf{C11}      & \textbf{Action Units (AUs) + Eye Gaze}                                                                    & 0.70& 0.54& 0.36& 0.43\\ \midrule
\textbf{C9}      & \textbf{Head Pose + Eye Gaze}                                                                             & 0.59& 0.55& 0.45& 0.48\\ \midrule
\textbf{C10}     & \textbf{Eye Gaze + Body Pose}                                                                             & 0.74& 0.64& 0.54& 0.57\\ \midrule
\textbf{C12}     & \textbf{\begin{tabular}[c]{@{}l@{}}Eye Gaze + Action Units (AUs) \\ + Head Pose\end{tabular}}             & 0.35& 0.38& 0.43& 0.43\\ \midrule
\textbf{C13}     & \textbf{\begin{tabular}[c]{@{}l@{}}Eye Gaze + Action Units (AUs)\\ + Body Pose\end{tabular}}              & 0.75& 0.68& 0.60& 0.61\\ \midrule
\textbf{C14}     & \textbf{\begin{tabular}[c]{@{}l@{}}Eye Gaze + Head pose \\ + Body Pose\end{tabular}}                      & 0.71& 0.57& 0.42& 0.48\\ \midrule
\textbf{C15}     & \textbf{\begin{tabular}[c]{@{}l@{}}Eye Gaze + Action Units (AUs)\\  + Head Pose + Body Pose\end{tabular}} & 0.69& 0.65& 0.65& 0.70\\ \bottomrule
\end{tabular}
\end{table}